\documentclass[aps, reprint,10pt, superscriptaddress]{revtex4-1}
\usepackage[latin1]{inputenc}
\usepackage{graphicx}
\usepackage{amsmath,amssymb,amstext,graphicx,bm,booktabs}
\usepackage{bm}
\usepackage{color,soul,xpatch}
\newcommand{\CM}{\ensuremath{C_\mathrm{M}} }
\newcommand{\CD}{\ensuremath{C_\mathrm{D}} }

\newcommand{\CT}{\ensuremath{C_\mathrm{T}} }
\newcommand{\CP}{\ensuremath{C_\mathrm{P}} }

\newcommand{\fE}{\ensuremath{f_\mathrm{E}} }

\newcommand{\fD}{\ensuremath{f_\mathrm{D}} }
\newcommand{\fC}{\ensuremath{f_\mathrm{C}} }
\newcommand{\fP}{\ensuremath{f_\mathrm{P}} }

\newcommand{\VDC}{\ensuremath{V_{\text{dc}}}}

\newcommand{\PC}{\ensuremath{P_{\text{C}}}}

\newcommand{\QE}{\ensuremath{Q_{\text{E}}}}

\newcommand{\Qij}{\ensuremath{Q_{\text{i,j}}}}

\newcommand{\go}{\ensuremath{g_0} }
\newcommand{\uzpm}{\ensuremath{u_\mathrm{ZP}}}

\newcommand{\CC}{\ensuremath{C_\mathrm{C}} }

\newcommand{\Qo}{\ensuremath{Q_\mathrm{0} }}

\newcommand{\VM}{\ensuremath{V_{\text{M}}}}

\newcommand{\fo}{\ensuremath{f_0} }

\newcommand{\ZO}{\ensuremath{Z_0} }
\newcommand{\Ztot}{\ensuremath{Z_\mathrm{tot}} }

\newcommand{\uzp}{\ensuremath{u_\mathrm{ZP}}}

\newcommand{\fij}{\ensuremath{f_{i,j}} }

\newcommand{\Cshunt}{\ensuremath{C_\mathrm{M}} }
\newcommand{\Cseries}{\ensuremath{C_\mathrm{D}} }
\newcommand{\CA}{\ensuremath{C_\mathrm{A}} }
\newcommand{\SV}{\ensuremath{S_{V}} }
\newcommand{\deltaf}{\ensuremath{\delta f}}

\begin{document}

\title{Radio-frequency optomechanical characterization of a silicon nitride drum}

\author{A.N.~Pearson}
\affiliation{Department of Materials, University of Oxford, Parks Road, Oxford, OX1 3PH, United Kingdom}

\author{K.~E.~Khosla}
\affiliation{Center for Engineered Quantum Systems, The School of Mathematics and Physics, The University of Queensland, St. Lucia, Queensland, 4072, Australia}
\affiliation{QOLS, Blackett Laboratory, Imperial College London, London, SW7 2AZ, United Kingdom}

\author{M.~Mergenthaler}
\affiliation{Department of Materials, University of Oxford, Parks Road, Oxford, OX1 3PH, United Kingdom}


\author{G.A.D.~Briggs}
\affiliation{Department of Materials, University of Oxford, Parks Road, Oxford, OX1 3PH, United Kingdom}

\author{E.A.~Laird}
\affiliation{Department of Physics, Lancaster University, Lancaster, LA1 4YB, United Kingdom}

\author{N.~Ares}
\email{natalia.ares@materials.ox.ac.uk}
\affiliation{Department of Materials, University of Oxford, Parks Road, Oxford, OX1 3PH, United Kingdom}

\begin{abstract}

On-chip actuation and readout of mechanical motion is key to characterize mechanical resonators and exploit them for new applications. 
We capacitively couple a silicon nitride membrane to an off resonant radio-frequency cavity formed by a lumped element circuit. Despite a low cavity quality factor ($\QE \approx 7.4$) and off resonant, room temperature operation, we are able to parametrize several mechanical modes and estimate their optomechanical coupling strengths. This enables real-time measurements of the membrane's driven motion and fast characterization without requiring a superconducting cavity, thereby eliminating the need for cryogenic cooling. Finally, we observe optomechanically induced transparency and absorption, crucial for a number of applications including sensitive metrology, ground state cooling of mechanical motion and slowing of light.         

\end{abstract}

\date{\today{}}
\maketitle

 \section*{Introduction}
Cavity optomechanics boasts a number of tools for investigating the interaction between radiation and mechanical motion and enables the characterization and development of highly sensitive devices~\cite{Aspelmeyer2014}. Silicon nitride membranes have been fabricated to exhibit very high tensile stress, resulting in high quality factors, and have been used for a number of applications including measurement of radiation pressure shot noise~\cite{Purdy2013}, optical squeezing~\cite{Purdy2013a}, bidirectional conversion between microwave and optical light~\cite{Andrews2014}, optical detection of radio waves~\cite{Bagci2014, Moaddel2018}, microkelvin cooling~\cite{Yuan2015} and cooling to the quantum ground state of motion~\cite{Nakamura2016, Peterson2016, Fink2016}.    

Radio-frequency (rf) cavities allow for sensitive mechanical readout on-chip~\cite{Lehnert2014, Bagci2014, Ares2016b, Brown2007, Faust2012}. We characterize a silicon nitride membrane at room temperature making use of an off-resonant rf cavity \cite{Brown2007, Faust2012}. In our approach, the use of lumped elements greatly simplifies the detection circuit in terms of fabrication and allows the integration on chip with the mechanical oscillator. Our circuit has a lower operation frequency than microwave cavities~\cite{Lehnert2014, Faust2012}, and allows for a larger readout bandwidth than previous works~\cite{Brown2007}. Also, our cavity allow us to inject noise, effectively increasing the mechanical mode temperature. We are able to detect several modes and extract the quality factor and cavity coupling strength for each of them. When the membrane is driven, we are able to resolve the membrane's motion in real time. We achieve a sensitivity of $0.4~\mathrm{pm}/\sqrt{\mathrm{Hz}}$.  A sensitivity of $4.4~\mathrm{pm}/\sqrt{\mathrm{Hz}}$ was reported in Ref.~\cite{Faust2012}, although it must be noted that these sensitivities cannot be easily compared, due to the much smaller size of their mechanical resonator. We observe optomechanically induced transparency (OMIT) and optomechanically induced absorption (OMIA) on-chip and deep in the unresolved sideband regime, allowing for the characterisation of the membrane's motion under radiation pressure. OMIT and OMIA are an unambiguous signature of the optomechanical interaction~\cite{Weis2010} and they can be used to slow or advance light~\cite{Safavi-Naeini2011}. OMIT has also been proposed as a means to achieve ground state cooling of mechanical motion in the unresolved sideband regime~\cite{Ojanen,Yong-Chun2015}.


\begin{figure*}
\includegraphics[scale=1]{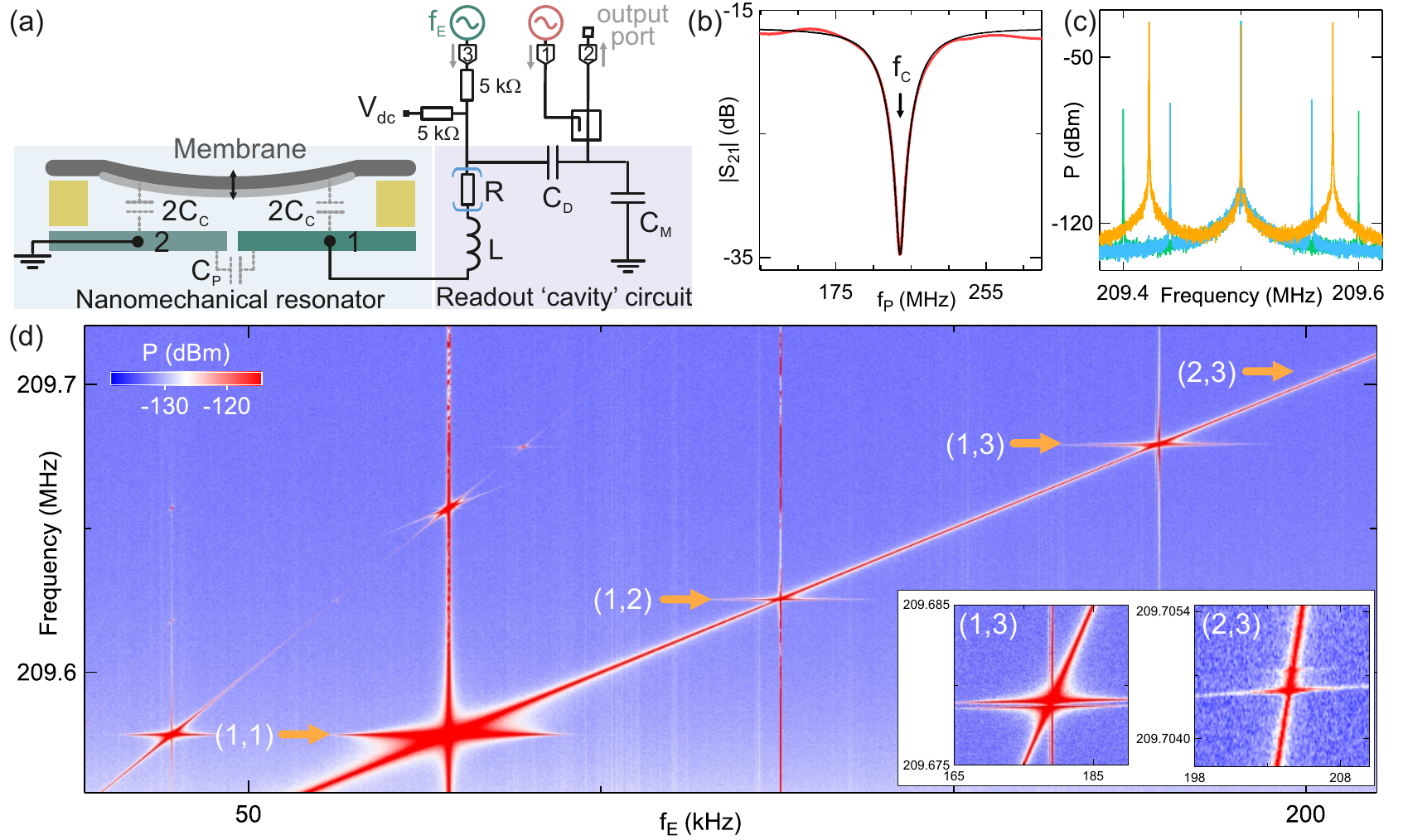}
\caption{\label{Fig1} (a)  Experimental setup. A metalized silicon nitride membrane is suspended over two metal electrodes. Electrode 2 is grounded and electrode 1 is connected to a radio-frequency tank circuit acting as a readout cavity. The cavity is formed from an inductor $L=223~$nH and 10~pF fixed capacitors $\CD$ and $\CM$. Parasitic capacitances contribute to $\CM$. The role of \CM is to improve on the impedance matching between the circuit and the 50 Ohm line. \CD controls the coupling to the cavity, i.e. the number of photons entering the cavity. Because these capacitors are larger than \CT, they do not strongly affect the cavity frequency. Parasitic losses in the circuit are parameterized by an effective resistance $R$. To probe the cavity a radio-frequency signal is injected at port 1, passed via a directional coupler, and after reflection received at port 2. The signal is measured using a vector network analyser or spectrum analyser. The membrane's motion is excited via port 3. Bias resistors allow a dc voltage $\VDC$ to be added to the membrane drive signal.  (b) Reflection as a function of frequency. The black curve is a fit to a circuit model. (c) Power spectrum of the signal at port 2 showing the carrier peak at the cavity resonance frequency, applied via port 1, and sidebands corresponding to a frequency modulated tone applied via port 3. The light blue and light green curves correspond to the power spectrum for an excitation $\fE$ below (60~kHz) and above (100~kHz) $\fo$ respectively. Sidebands appear due to non-linearities in the circuit. The orange curve corresponds to $\fE\sim\fo$. In this case, the sidebands are larger. (d) Higher frequency sidebands in (c) as a function of $\fE$. When $\fE$ is close to a mechanical mode of the membrane, a mechanical sideband appears. The four visible mechanical modes are indicated with arrows. Because higher modes produce fainter features, they are not displayed in this graph. Note that the colour scale displays a range of only 25dBm in order to reveal the weaker features. Insets: zoom-in of the two highest frequency modes displayed. We observe a splitting of the mechanical feature, indicating nearly degenerate modes.}
\end{figure*}

\section*{Experiment}
Our device consists of a high-stress silicon nitride membrane which is 50~nm thick; it has an area of 1.5~mm$\times$1.5~mm and $90\%$ of this area is metalized with 20~nm of Al. We suspend this membrane over two Cr/Au electrodes patterned on a silicon chip. A dc voltage $\VDC=15$~V is applied to electrode 1, with electrode 2 grounded. Measurements were performed at room temperature and at approximately $10^{-6}$~mbar. For optomechanical readout and control, electrode 1 is connected to an effective rf cavity. The cavity is realised using an on-chip inductor $L$ and capacitors $\CD$ and $\CM$ mounted on the sample holder~\cite{Ares2016}, in addition to the capacitance formed by the membrane $\CC$. The circuit behaves similarly to a simple LC resonator with total capacitance  $\CT = \CC + \CP$, where $\CP$ accounts for the capacitance between the two sides of the antenna and other parasitic capacitances. This circuit acts as a cavity and can be driven by injecting an rf signal to the input (port 1) via a directional coupler in order to induce an optomechanical interaction between the cavity signal and the mechanical motion.  In addition port 3 allows injection of an ac signal to directly excite the membrane's motion. The entire setup forms a three-terminal circuit with input ports 1 and 3 and output port 2 (Fig.\ref{Fig1}(a)). We used a vector network analyzer to measure scattering parameters and a spectrum analyzer to measure power spectra. Figure~\ref{Fig1}(b) shows the scattering parameter ($|\mathrm{S}_{21}|$) as a function of cavity probe frequency $\fP$. The cavity resonance is evident in reflection as a minimum in $|\mathrm{S}_{21}|$ with quality factor $\QE \approx 7.4$.


The dependence of the capacitor formed between the electrodes and the metalized membrane $\CC$ on the mechanical displacement $u$ leads to coupling between the cavity and the mechanical motion. This coupling is given by $\frac{d \fC}{du} \approx 1/(4\pi\sqrt{L\CT^3})\frac{\partial \CC}{\partial u}$, where $\fC$ is the cavity resonance frequency, $\CT$ the total circuit capacitance and $u$ the membrane's displacement from its equilibrium position~\cite{Supp}. The coupling causes phase shifts of the cavity reflection, allowing us to monitor the membrane's motion in the unresolved sideband limit. The single-photon coupling strength, which measures the interaction between a single photon and a single phonon, is therefore,
\begin{equation}
\frac{\go}{2\pi} \approx\frac{1}{4\pi\sqrt{L\CT^3}}\frac{\partial \CC}{\partial u}\uzpm,
\label{g0}
\end{equation} 
 where $\uzpm$ is the amplitude of the membrane's zero-point motion.

Using a simple circuit model~\cite{Supp}, we fit $|\mathrm{S}_{21}|$ and extract $\CC\approx1.6$~pF. Within the parallel plate capacitor approximation, $\CC\sim\frac{\epsilon_0 a}{d}$, where $\epsilon_0$ is the vacuum permittivity, $a$ is the metallised area of the membrane and $d$ is the membrane-electrode gap. From this expression, we extract $d\sim 9~\mu$m. 

 \section*{Results}

To find the mechanical resonances, we drove the cavity on resonance ($\fC \approx 209.2$~MHz) with input power $\PC=5$~dBm at port 1. Meanwhile, we excited the membrane's motion via port 3, using a sinusoidal signal at frequency $\fE$ and with amplitude $\VM=3.6$~V$_\mathrm{rms}$ at electrode 1. In order for the mechanical response to appear broader in the frequency spectrum, facilitating the detection of the mechanical modes, the excitation frequency at port 3 was modulated with a white noise pattern with a deviation of 200~Hz. The power spectrum $P$ of the reflected signal at port 2 shows sidebands at $\fC \pm \fE$  due to non-linearities of the rf circuit giving rise to frequency mixing. The mechanical signal appears when $\fE$ is close to a mechanical resonance, and is evident as a pronounced increase in sideband amplitude and width (Fig.~\ref{Fig1}(c)).

Figure~\ref{Fig1}(d) shows the sideband at $\fC + \fE$ as a function of $\fE$. The fundamental mode frequency $f_{1,1}$, which we will call $\fo$, is $\sim77.9$~kHz, giving an unresolved sideband ratio of $2\pi\fo/\kappa\sim3\times10^{-3}$, with $\kappa = 2\pi \fC/\QE$ the cavity linewidth. As well as the fundamental mode, we observe less strong harmonics $\fij$ near the expected frequencies. The expected frequencies for higher harmonics theoretically satisfy the ratios $\fij/\fo=\frac{1}{\sqrt{2}}\sqrt{i^2+j^2}$ for integers $i$ and $j$ with symmetric roles as expected for a square membrane. Two of the sidebands are double peaked, evidencing nearly degenerate mechanical modes (insets Fig.~\ref{Fig1}(d)). The broken degeneracy could be due to imperfections in how the membrane was fabricated and fixed in place or uneven binding/deposition of the Al layer. 

\begin{figure}[h]

\includegraphics[scale=1]{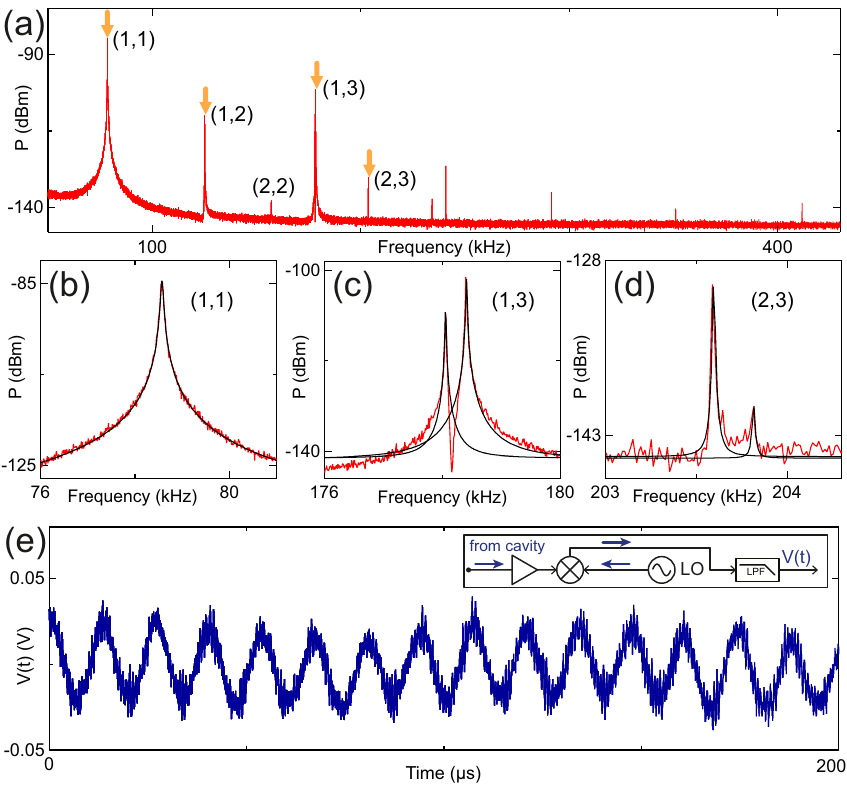}
\caption{\label{Fig2}(a) Power spectrum as a function of frequency for the higher-frequency sidebands.  A tone at $\fC$ is applied with input power $\PC= 5$~ dBm and white noise is injected via port 3 at $\VM=2.7$~V$_{\mathrm{rms}}$ at electrode 1. We observe several mechanical modes. The four mechanical modes visible in Fig.~\ref{Fig1}(d) are indicated with arrows. The mode at $\sim157$~kHz, labelled (2,2), is not visible in Fig.~\ref{Fig1}(d) due to a higher noise floor than in Fig.~\ref{Fig2}(a), which arises from a higher resolution bandwidth. Higher modes are not labelled because the deviation from theoretical ratios makes their identification ambiguous. (b-d) Zoom in showing the fundamental mode, and selected nearly degenerate modes. The black curves are Lorentzian fits. e) Demodulated output signal as a function of time for the fundamental mode. The demodulation circuit is shown in the inset. (LO: local oscillator; LPF: low-pass filter). Input power is \PC = 12 dBm at \fC and the white noise is \VM = 3.6 V$_\mathrm{rms}$ at electrode 1.  
}

\end{figure}

 
The entire set of mechanical resonances can be observed in a single measured power spectrum by driving the cavity at $\fC$ and injecting white noise via port 3. White noise excites the motion of the membrane at all frequencies, which is equivalent to raising the effective mechanical temperature. In this way, the root mean square (rms) displacement is increased, thereby facilitating the detection of mechanical modes. The noise signal has a bandwidth of 1~MHz, larger than the spectral width of the mechanical modes, and an amplitude $\VM$ = 2.7 V$_\mathrm{rms}$. Figure~\ref{Fig2}(a) shows the mechanical sidebands at $\fC + \fij$. To distinguish mechanical sidebands from other parasitic signals, we increase $\PC$ until we observe a frequency shift~\cite{Supp}. The fundamental mode of the membrane is at $\fo=78.573\pm0.002$~kHz. We fit each mechanical sideband with a Lorentzian (Figs.~\ref{Fig2}(b-d)). As in Fig.~\ref{Fig1}(d), we observe double peaked sidebands (Figs.~\ref{Fig2}(c,d)).

The broad cavity bandwidth allowed us to measure the actuated membrane's motion in real time. We excite the membrane with white noise whilst driving the cavity at \fC. In order to record the motion in real time, the cavity output signal is mixed with a local oscillator. The output signal (Fig.~\ref{Fig2}(e)) shows clear sinusoidal oscillations evidencing the membrane's motion.

\begin{figure}[h]
\includegraphics[scale=1]{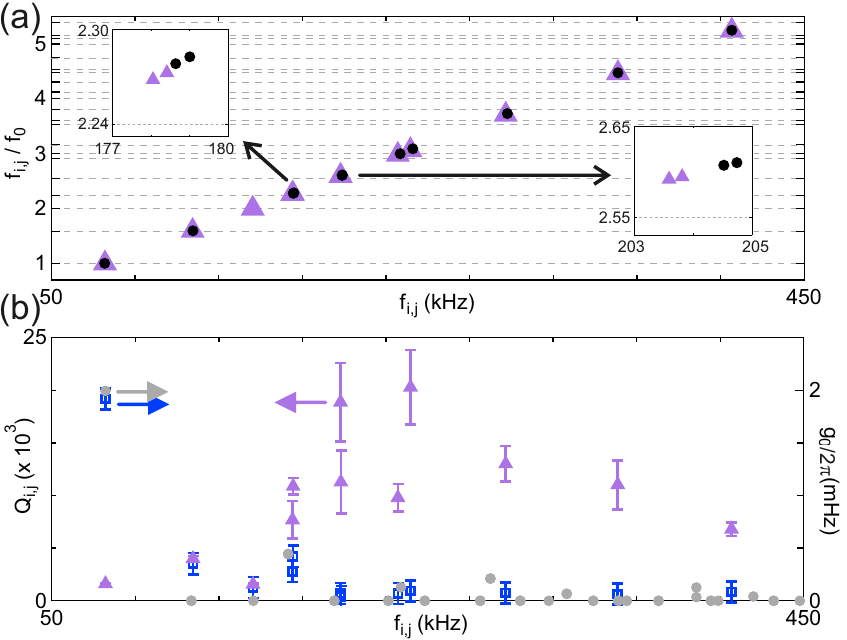}
\caption{(a) Frequency ratios of the mechanical resonances extracted from Fig.~\ref{Fig1}(d) (circles) and Fig.~\ref{Fig2}(a) (triangles). Grid lines indicate theoretical ratios. Errors in frequency and frequency ratios are smaller than symbols. Insets: zoom-in showing selected nearly degenerate modes. (b) Triangles: Quality factors extracted from Lorentzian fits as in Fig.~\ref{Fig2} for each mechanical mode. Error bars were obtained by combining fit results for different values of $\VM$. Squares: Single-photon coupling strength calculated for each mechanical mode. Circles: Single-photon coupling strength estimated from a parallel capacitor model, corrected for different mode profiles, for $\fij$ obtained from theoretical ratios. 
\label{Fig3}
}
\end{figure}
\begin{figure}
\includegraphics[scale=1]{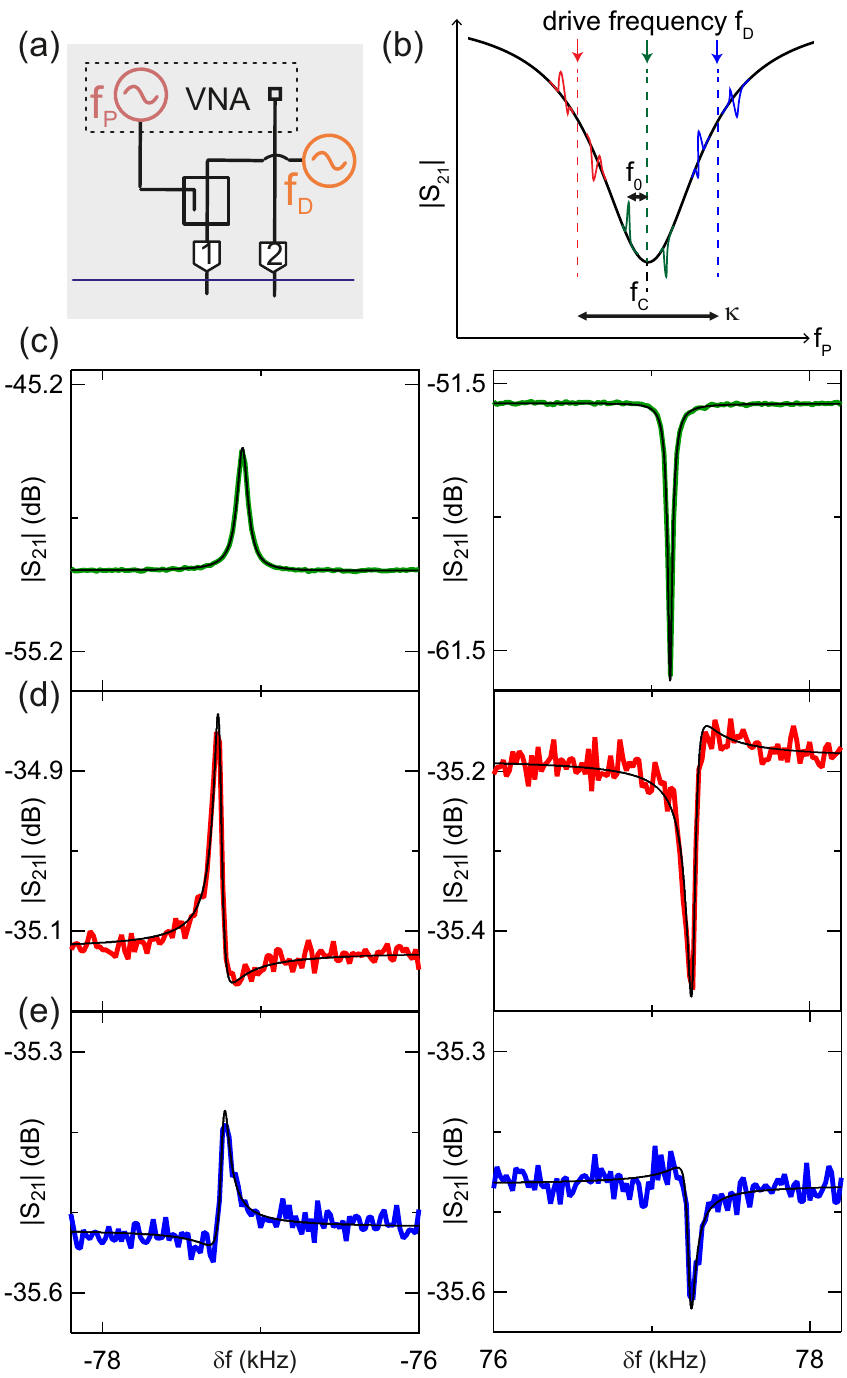}
\caption{(a) Schematic of signal injection and detection for OMIT/OMIA measurements. A tone with frequency $\fD$ and a tone with frequency $\fP$ are injected through a directional coupler via port 1. We monitor $|\mathrm{S}_{21}|$ with a vector network analyzer. (b) Schematic showing each of the drive frequencies injected and a cartoon of the mechanical sidebands appearing at $\fD\pm\fo$.
Drive frequencies were $\fD\sim\fC$ (green), $\fD\sim\fC-\kappa/2$ (red) and $\fD\sim\fC+\kappa/2$ (blue). The corresponding mechanical sidebands are shown in panels (c), (d) and (e) respectively. (c-e) Measured $|\mathrm{S}_{21}|$ as a function of $\delta f$ showing mechanical sidebands at $\delta f\sim-\fo$ (left panel) and $\delta f\sim\fo$ (right panel) for each of the values of $\fD$ considered in (b). At port 1, the drive power was 5 dBm and the probe power -27.5 dBm. Black curves are a fit to the data \cite{Supp}.
\label{Fig4}
}
\end{figure}

We plot $\fij/\fo$ as a function of $\fij$  in Fig.~\ref{Fig3}(a) for all mechanical resonances observed in Fig.~\ref{Fig1}(d) and Fig.~\ref{Fig2}(a). Lower frequency modes show better agreement with theoretical ratios than higher frequency modes. From the Lorentzian fits of each mechanical sideband, we extract the mechanical quality factors $\Qij$ and single photon coupling strengths $\go$ (Fig.~\ref{Fig3}(b)). These values of $\Qij$, measured in the spectral response, are sensitive both to dissipation and dephasing~\cite{Schneider2014}. The fundamental mode has a quality factor $\Qo=(1.6\pm0.1)\times10^3$ and the highest quality factor measured was $(20\pm4)\times10^3$ for the mode at 241~kHz. Different modes and even nearly degenerate mechanical modes have significant differences in their values of $\Qij$, as previously reported~\cite{Zwickl2008,Yuan2015a}. The values of $\Qij$ vary slightly as a function of $\VM$ ranging from 0.1 to 3.6~V$_{\mathrm{rms}}$, but they do not show a specific trend~\cite{Supp}. The error bars correspond to this deviation in the values of $\Qij$.

We extract $\go$ for each mechanical mode from the effective thermomechanical power, i.e. from the area of the corresponding sideband in Fig.~\ref{Fig3}(b)~\cite{Supp}. For the first mechanical mode, we obtain $\go/2\pi=1.9\pm0.1$~mHz. The cavity drive was 5~dBm at port 1, yielding a multiphoton coupling strength $g/2\pi = \sqrt{n_\mathrm{C}}\go/2\pi\sim4.4$~kHz, where $n_\mathrm{C}$ is the mean cavity photon number. As expected, the first mode couples more strongly than higher frequency modes, due to its mode profile and larger zero point motion~\cite{Supp}.  

These values of $\go$ can be compared with the predictions of Eq.~\ref{g0}. Taking $\CT$ and $d$ from the circuit model, and using that $\frac{\partial \CC}{\partial u}\sim\frac{\epsilon_0 a}{d^2}$ for a parallel-plate capacitor (with a known prefactor to account for the mechanical mode profile of the membrane), gives a coupling strength $g_0/2\pi \approx 2$~mHz for the fundamental mode [18].  The estimated values of $\go$ are similar to the ones extracted from the sideband powers.

We can estimate the vibrational amplitude sensitivity $S_u$ given an amplifier voltage sensitivity $S_v=\sqrt{k_\mathrm{B} T_\mathrm{N} Z_0}$, where $k_\mathrm{B}$ is the Boltzmann constant, $Z_0\equiv50~\Omega$ and $T_\mathrm{N}\sim293$~K the system noise temperature. We can write $S_u=S_v/(\frac{\partial |S_{21}|}{\partial \CT} \frac{\partial \CC}{\partial u} V_\mathrm{in} )$ where $ V_\mathrm{in}$ is the voltage at electrode 1 corresponding to $\PC$. Taking $\frac{\partial |S_{21}|}{\partial \CT}$ from the circuit model and $\frac{\partial \CC}{\partial u}\sim1.7\times10^{-7}$~F/m extracted from the sideband's area corresponding to the fundamental mode~\cite{Supp}, we obtain $S_u=0.4~\mathrm{pm}/\sqrt{\mathrm{Hz}}$. This value is comparable to that obtained for a suspended carbon nanotube device at cryogenic temperatures~\cite{Ares2016b}.

Finally, we measure optomechanically induced transparency (OMIT) and absorption (OMIA), which are signatures of optomechanical coupling and demonstrate that the membrane's motion can be actuated by radiation-pressure alone. OMIT and OMIA are characterized by the emergence of a transparency  or absorption window in $|\mathrm{S}_{21}|$ when a strong drive tone ($\fD$) and a weak probe tone ($\fP$) are injected into the cavity, and the frequency difference between these tones $\delta f=\fP-\fD$ coincides with the frequency of a mechanical mode~\cite{Weis2010}. When this condition is met, the beat between the drive and the probe field excites the membrane's motion and the destructive (constructive) interference of excitation pathways for the intracavity probe field results in a transparency (absorption) window in $|\mathrm{S}_{21}|$.

To show optomechanical control, we measured OMIT and OMIA by injecting a strong tone at frequency $\fD$ and a weak tone at frequency $\fP$ through a directional coupler at port 1 (Fig.\ref{Fig4}(a)). We injected three different drive frequencies (Fig.\ref{Fig4}(b)); $\fD\sim\fC-\kappa/2$ (red detuned), $\fD\sim\fC$ (resonant) and $\fD\sim\fC+\kappa/2$ (blue detuned).
When $\fD\sim\fC$, a peak (dip) is observed at $\fD-(+)\fo$  (Fig.~\ref{Fig4}(c)). When $\fD\sim\fC\pm\kappa/2$, we observe Fano-like features at $\fD\pm\fo$ (Fig.~\ref{Fig4}(d,e)). We do not expect complete extinction or revival of the reflected signal as the system is well within the unresolved sideband limit.

We fitted OMIT and OMIA features by modelling the transmission of the probe field (for details see \cite{Supp}). From the fit of the spectral features, we extract $\fo=77.2\pm0.1$~kHz, $\Qo=(1.2\pm0.1)\times10^3$ and  $\go/2\pi=2.3\pm0.3$~mHz. Error bars were obtained by combining fit results of the six curves in Fig.~\ref{Fig4}(c-e). The values obtained for $\fo$ and $\Qo$ are similar to those extracted from the Lorentzian fits as in Fig.~\ref{Fig2}(b). The value of $\go$ is in agreement with the extracted from Fig.~\ref{Fig2}(a) and estimated from the parallel plate capacitor approximation. 

\section*{Discussion}
To conclude, we have characterized several modes of a silicon nitride membrane with an off-resonant rf circuit at room temperature, deeper in the unresolved sideband regime than previously explored.  Our cavity allows for the injection of noise to actuate the motion of the membrane and effectively increase its mechanical mode temperature. We achieve a sensitivity of $0.4~\mathrm{pm}/\sqrt{\mathrm{Hz}}$, a tenfold improvement to that reported in Ref.~\cite{Faust2012} although the smaller size of their mechanical resonator makes direct comparison difficult. Our results show that our on-chip platform can be used for membrane actuation and characterization. The readout circuit operates at a convenient frequency and does not require the cavity to be tuned into resonance with the membrane as in other approaches. It therefore has applications in mechanical sensing and microwave-to-optical conversion. Thanks to the large bandwidth of the cavity we have also measured the actuated membrane's motion in real time.  Finally we have observed OMIT and OMIA, from which we obtained a separate measure of the frequency, quality factor and coupling strength of the fundamental mode. \\


\section*{Data Availability}
The data analysed during the current study are available from the corresponding author on reasonable request. 

\section*{Acknowledgements}

We acknowledge discussions with G. J. Milburn and support from EPSRC (EP/J015067/1), the Royal Society, the Royal Academy of Engineering and Templeton World Charity Foundation. The opinions expressed in this publication are those of the authors and do not necessarily reflect the views of Templeton World Charity Foundation. KK acknowledges travel funding from the Australian Research Council (CE110001013), and thanks the Oxford Materials Department for hospitality during the initial stages of this work.

\section*{Author contributions statement}

A.N.P. and N.A. performed the experiment, analysed the data and prepared the manuscript with contributions from E.A.L. K.E.K. contributed to the theory, M.M. helped with device preparation. N.A. conceived the experiment. A.N.P., K.E.K., M.M., E.A.L., G.A.D.B. and N.A. discussed results and commented on the manuscript.

\section*{Additional information}

\textbf{Competing interests} The authors declare no competing interests.

%



\onecolumngrid

\vfill

\newpage

\newcommand{\beginsupplement}{%
        \setcounter{table}{0}
        \renewcommand{\thetable}{S\arabic{table}}%
        \setcounter{figure}{0}
        \renewcommand{\thefigure}{S\arabic{figure}}%
             \setcounter{equation}{0}
        \renewcommand{\theequation}{S\arabic{equation}}%
                     \setcounter{page}{1}
     }
        
      \parbox[c]{\textwidth}{\protect \centering \Large \MakeUppercase{Supplementary Information}}
\rule{\textwidth}{1pt}
\vspace{11pt}
\onecolumngrid

\beginsupplement
\title{Radio-frequency optomechanical characterization of a silicon nitride drum}
\maketitle

\author{A.~Pearson}
\affiliation{Department of Materials, University of Oxford, Parks Road, Oxford OX1 3PH, United Kingdom}

\author{K.~E.~Khosla}
\affiliation{Center for Engineered Quantum Systems, The University of Queensland, Brisbane, Queensland 4067, Australia}
\affiliation{School of Mathematics and Physics, The University of Queensland, Brisbane, Queensland 4067, Australia}
\affiliation{QOLS, Blackett Laboratory, Imperial College London, London SW7 2AZ, United Kingdom}

\author{M.~Mergenthaler}
\affiliation{Department of Materials, University of Oxford, Parks Road, Oxford OX1 3PH, United Kingdom}


\author{G.A.D.~Briggs}
\affiliation{Department of Materials, University of Oxford, Parks Road, Oxford OX1 3PH, United Kingdom}

\author{E.A.~Laird}
\affiliation{Department of Physics, Lancaster University, Lancaster, LA1 4YB, United Kingdom}

\author{N.~Ares}
\affiliation{Department of Materials, University of Oxford, Parks Road, Oxford OX1 3PH, United Kingdom}


\section{Cavity characterisation}
\subsection{Fit of the cavity transmission}
\label{sec:sectionI}
The response of the cavity, Fig.~1 (b) of the main text, was characterized by fitting the measured transmission to the following expression:
\begin{equation}
S_{21}(\fP)=\alpha e^{i\phi}+(1-\alpha)(1-\frac{\kappa_e}{-2\pi i(\fP-\fC)+\kappa/2})
\label{eq:s21}
\end{equation} 
where $\alpha$ is the isolation of the directional coupler and $\phi$ a phase~\cite{Singh2014}. The cavity resonance frequency is $\fC$ , $\kappa_e$ is the external dissipation rate and $\kappa$ the total dissipation rate of the cavity.

From a fit with a fixed value of $\alpha=0.001$ from the directional coupler data sheet and correcting for an insertion loss of -16.4~dB, we extract $\phi=(0.9\pm0.1)\pi$~rad, $\fC=209.23\pm0.01~$MHz, $\kappa/(2\pi)=28.20\pm0.01~$MHz and $\kappa_e/(2\pi)=15.80\pm0.01~$MHz.

\subsection{Circuit simulation}
\label{sec:sectionII}

\begin{figure}[h]
\includegraphics[scale = 1]{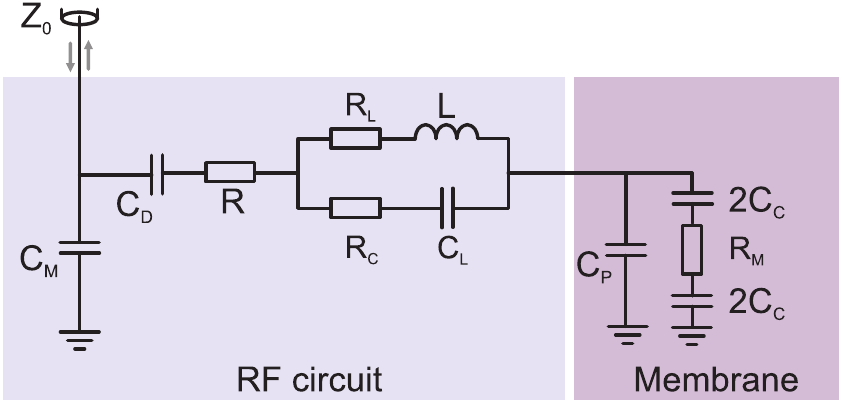}
\caption{\label{S1}
Circuit model. Capacitors $\Cshunt$ and $\Cseries$  are taken as simple lumped elements, including any parasitic capacitances in parallel. Elements  $R_{\text{L}}$, $R_{\text{C}}$ and $C_{\text{L}}$ model parasitic contributions to the impedance of the inductor $L$. The effective resistance $R$ models other losses in the circuit. The membrane is modelled by the combination of $R_{\text{M}}$ and $C_{\text{C}}$, the capacitance between the metallised area of the membrane and both antenna electrodes. To each electrode, the membrane-electrode capacitance is $2C_{\text{C}}$, which when summed together in series gives $C_{\text{C}}$.
}
\end{figure}

To extract circuit parameters the cavity response can also be fit using the circuit model in Fig.~\ref{S1}, as shown in Fig.~1 (b). The capacitors $\Cseries$ and $\Cshunt$ are taken as simple lumped elements. We consider the capacitor formed between the electrodes and the metalized membrane, $\CC$, formed by  two electrode-membrane capacitors of value $2\CC$, as well as the capacitance between antenna electrodes, $\CA$. The inductor is modeled as a network of elements as shown, which simulate its self-resonances and losses. The membrane has a resistance $R_M$ and other losses in the circuit are modeled by an effective resistance $R$. 

The reflection coefficient $\Gamma$ is then equal to
\begin{equation}
\Gamma(\fP) = \frac{\Ztot(\fP)-\ZO}{\Ztot(\fP)+\ZO},
\label{eq:gammaZtot}
\end{equation} 
where $\Ztot$ is the total impedance from the circuit's input port and $Z_0=50~\Omega$ is the line impedance. We relate the measured transmission $S_{21}$ to $\Gamma$ by assuming a constant overall insertion loss $A$, incorporating  attenuation in the lines, the coupling of the directional coupler, and the gain of the amplifier, such that 
\begin{equation}
|S_{21}(\fP)|=A|\Gamma(\fP)|.
\label{eq:s21b}
\end{equation} 

Fitting to Eq.~(\ref{eq:s21}), we take $\Cseries=10$~pF from the known component value, and $L=223$~nH, $R_{\text{L}}=3.15 \times 10^{-4}~\Omega \times \sqrt{\fP~[\mathrm{Hz}]}$, R$_{\text{C}}=25~\Omega$ and C$_{\text{L}}=0.082$~pF from the datasheet of the inductor. For the resistance of the aluminium film, we estimate $R_\mathrm{M}$ = 7.5 $\Omega$ using the resistivity of aluminium (15 $\mu \Omega$ cm) and the known film thickness (20 nm) \cite{Lacy2011}. We have estimated $1.4$~pF from a COMSOL model of the antenna electrodes and we have added a parasitic capacitance of $0.7$~pF based on previous work~\cite{Ares2016supp} making a total parasitic capacitance $\CP$ of $2.1$~pF. 
Fit parameters are then $A$, $\Cshunt$, $R$, and $\CC$. From the fit we obtain $A=-16.657\pm0.001$~dB, $\Cshunt=20.93\pm0.01$~pF, $R=15.87\pm0.01~\Omega$ and $\CC=1.6432\pm0.0001$~pF. 

The cavity coupling to the membrane displacement is given by $\frac{d \fC}{du}= \frac{\partial\fC}{\partial \CT}\frac{\partial \CC}{\partial u}$ and $\frac{\partial \fC}{\partial \CT} \approx 1/(4\pi\sqrt{L\CT^3})$ where $\CT\equiv\CP+\CA+\CC$. Although this approximation applies strictly only for a simple LC resonator, we confirmed numerically that this procedure gives a good approximation for $\frac{\partial \fC}{\partial \CT}$.

\section{Mechanical Characterization} 
\subsection{Mechanical quality factors}
\label{sec:Qnm}
\begin{figure}[h]
\includegraphics[scale =1]{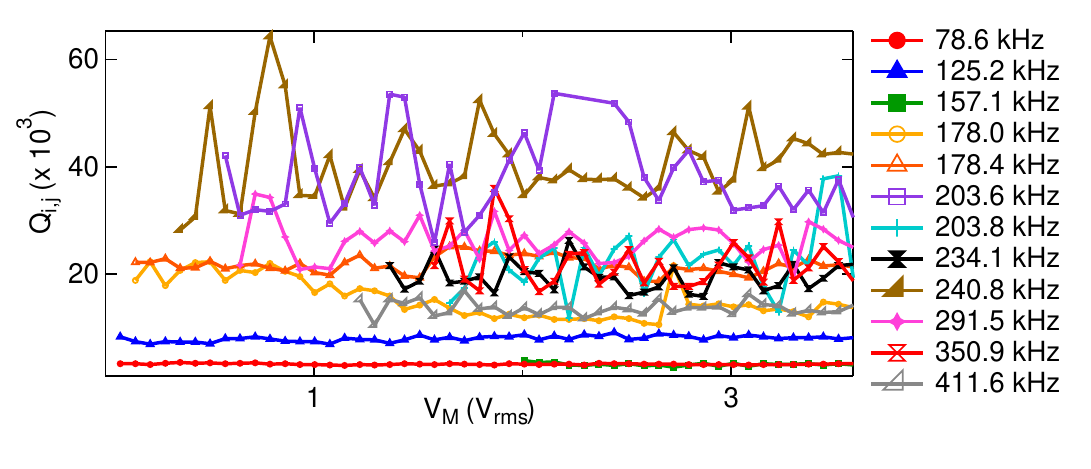}
\caption{\label{S2}
Extracted $\Qij$ as a function of $\VM$ for each mechanical mode observed. 
}
\end{figure}

From Lorentzian fits as in Fig. 2(b-d) of the main text we can extract $\Qij$. We extracted $\Qij$ for different values of $\VM$ (Fig.~\ref{S2}). At low $\VM$, the mechanical sidebands became fainter, and some modes could not be reliably fitted. The values of $\Qij$ do not show a trend as a function of $\VM$. The error bars in $\Qij$ in Fig. 3(b) of the main text were obtained, for each mechanical mode, by combining $\Qij$ for different values of $\VM$.

\begin{figure}[h]
\includegraphics[scale = 1.3]{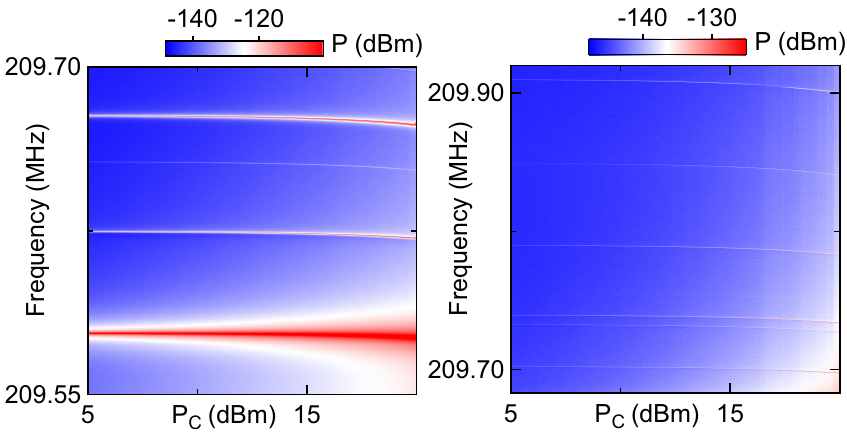}
\caption{\label{S4}
Power spectrum as a function of frequency and $\PC$ for $\VM =2.7~\mathrm{V_{rms}}$. The cavity drive is at 209.5~MHz. Several mechanical sidebands can be observed. The first few mechanical modes are displayed in (a) and the higher frequency mechanical modes are displayed in (b). As $\PC$ increases, the mechanical sidebands are brighter and their frequency shifts to lower values for $\PC\gtrsim15$~dBm.
}
\end{figure}

\subsection{Mechanical sidebands as a function of cavity drive}
In order to distinguish the mechanical sidebands in Fig.~2(a) from parasitic resonances we measure them as a function of $\PC$ (Fig.~\ref{S4}), as the frequency of the mechanical modes decreases for the highest values of $\PC$. This might have to do with heating of the membrane surface and thereby a decrease in its tension. From the circuit model we can calculate the power dissipated in the membrane. For $\PC=15$~dBm, the power dissipated is $\sim4~\mu$W. We estimate the maximum temperature increase by assuming that all the dissipated power is emitted as thermal radiation. Taking the emissivity of aluminum as 0.09 and applying the Stefan-Boltzmann law leads to an estimated temperature increase of $4.5$~K. For $\PC=20$~dBm, the temperature difference is $\sim13.5$~K.
A similar calculation confirms that the injected noise does not change the temperature of the membrane significantly.

\section{Electromechanical coupling}
\subsection{Extraction of $\go$ from mechanical sidebands}
From the area below the sidebands in Fig.~2(a) of the main text, we extracted the values of $\go$ plotted in Fig.~3(b) of the main text. In this Section we will derive the expression relating $g_0$ to the effective thermomechanical power ($P_\mathrm{side}$) extracted from this area. We start with $g_0/2\pi =\frac{d \fC}{du} u_\mathrm{ZP}= \frac{\partial\fC}{\partial \CT}\frac{\partial \CC}{\partial u}u_\mathrm{ZP}$. As discussed in Section~\ref{sec:sectionII} $\frac{\partial\fC}{\partial \CT}\approx (4\pi\sqrt{L\CT^3})^{-1}$ giving 

\begin{equation}
g_0/2\pi\approx(4\pi\sqrt{L\CT^3})^{-1} \frac{\partial \CC}{\partial u}u_\mathrm{ZP},
\label{eq:go}
\end{equation} 
where $u_\mathrm{ZP}=\sqrt{\hbar/(4\pi m\fij)}$ is the zero point motion of the membrane with effective mass $m\sim4.5\times10^{-10}$~kg.
We choose to normalize the mode eigenfunctions such that the effective mass equals the mass of the suspended segment for all modes~\cite{Poot2012}. The value of $L$ is known, $\CT$ is obtained from the circuit model fit and $u_\mathrm{ZP}$ can be estimated from the values of $\fij$ extracted from the Lorentzian fit of the mechanical sidebands. To estimate $\frac{\partial \CC}{\partial u}$, we write $P_\mathrm{side}$, which obeys

\begin{align}
P_\mathrm{side} =  \tilde{P}_\mathrm{C}~ 2 \bar{n}_\mathrm{i,j} \left(\frac{\kappa_e}{\kappa}\right)^2 \frac{g_0^2}{(\frac{\kappa}{2})^2 + (2\pi\fij)^2},
\label{eq:Pside}
\end{align}
where $\tilde{P}_\mathrm{C}$ is the cavity drive ${P}_\mathrm{C}$ having taken into account the overall insertion loss and $\bar{n}_\mathrm{i,j}$ is the phonon occupancy of the mode~\cite{Yuan2015supp}. Replacing $g_{0}^2$ with Eq.~(\ref{eq:go}),

\begin{align}
P_\mathrm{side}\approx \tilde{P}_\mathrm{C}~2\bar{n}_\mathrm{i,j}\left(\frac{\kappa_e}{\kappa}\right)^2 \frac{\left( \frac{\partial \CC}{\partial u}\right)^2 u_\mathrm{ZP}^2}{((\frac{\kappa}{2})^2 + (2\pi\fij)^2)(2\sqrt{L\CT^3})^{2}}.
\label{eq:Pside2}
\end{align}

For the measurements in Fig. 2 of the main text $\PC=5$~dBm at port 1. The values of $\kappa$, $\kappa_e$ and $A$ are obtained from the cavity characterization (Section~\ref{sec:sectionI}) and the circuit model fit (Section ~\ref{sec:sectionII}). 

We now write $\bar{n}_\mathrm{i,j}$,

\begin{equation}
\bar{n}_\mathrm{i,j} = \frac{m~(2\pi\fij)^2\langle\delta u_{\mathrm{i,j}}^2\rangle}{\hbar~(2\pi\fij)},
\label{eq:nm0}
\end{equation} 
where $\delta u_{\mathrm{i,j}}$ is the membrane displacement from its equilibrium position. In order to estimate the rms value $\langle \delta u^2_{\mathrm{i,j}} \rangle$, we write the effective electromechanical force on the membrane,

\begin{equation}
 F(t)=\frac{1}{2}V^2(t)\frac{\partial C_C}{\partial u},
\end{equation}
where $V(t)=\VDC+\delta V(t)$. The time-independent part $\VDC=15$~V is much larger than the time-dependent part $\delta V(t)$. The time-dependent part of $F(t)$ is to lowest order 

\begin{equation}
\delta F(t)=\VDC\frac{\partial C_C}{\partial u}\delta V(t),
\label{eq:F_el}
\end{equation}
where we assumed this electronic fluctuating force is much larger than the thermal noise.

In the frequency domain, the displacement is

\begin{equation}
\delta u_{\mathrm{i,j}}(f) = \chi_{\mathrm{i,j}}(f)\delta F(f),
\end{equation}
where the mechanical susceptibility is $\chi_{\mathrm{i,j}}(f)=\frac{1}{4\pi^2m}[\fij^2-f^2+if\fij/\Qij]^{-1}$~\cite{Aspelmeyer2014supp,Lehnert2014supp}. The values of $\Qij$ can be extracted from the Lorentzian fit of the mechanical sidebands (Section~\ref{sec:Qnm}).
We calculate $[\delta u_{\mathrm{i,j}}^2 ]_\mathrm{rms}$,

\begin{equation}
[\delta u_{\mathrm{i,j}}^2]_\mathrm{rms}= \VDC^2  \left(\frac{\partial \CC}{\partial u}\right)^2\int\limits_{-\infty}^{\infty}\int\limits_{-\infty}^{\infty} d f  d f' \langle \delta V(f)  \delta V(f') \rangle \langle \chi_{\mathrm{i,j}}(f) \chi_{\mathrm{i,j}}(f') \rangle,
\end{equation}
where $\chi_{\mathrm{i,j}}(f)$ and $\delta V(f)$ are uncorrelated. The fluctuating voltage $\delta V(f)$ is assumed to be well approximated by white noise over the frequency range of interest, with $ \langle \delta V(f)  \delta V(f') \rangle =\frac{\SV^2}{2}~\delta(f+f')$, where $\SV$ is the single-sided white noise power spectrum of the driving voltage $V$. 
We obtain

\begin{equation}
[\delta u_{\mathrm{i,j}}^2]_\mathrm{rms}= \VDC^2  \left(\frac{\partial \CC}{\partial u}\right)^2 \left(\int\limits_{0}^{\infty} df |\chi_{\mathrm{i,j}}(f)|^2\right) \frac{S_V^2}{2},
\label{eq:displacement}
\end{equation}
given that $\chi_{\mathrm{i,j}}(f)^*=\chi_{\mathrm{i,j}}(-f)$.

We can now rewrite Eq.~\ref{eq:nm0},

\begin{equation}
\bar{n}_\mathrm{i,j}= m (2\pi\fij/\hbar)~ \VDC^2 \left(\frac{\partial \CC}{\partial u}\right)^2 \frac{\SV^2}{2} \int df |\chi_{\mathrm{i,j}}(f)|^2.
\label{eq:nm}
\end{equation}

Replacing Eq.~\ref{eq:nm} in Eq.~\ref{eq:Pside}, we obtain the following expression,

\begin{equation} 
\frac{\partial \CC}{\partial u}=\sqrt[4]{\frac{P_\mathrm{side}}{\tilde{P}_\mathrm{C}} \left(\frac{\kappa}{\kappa_e}\right)^2 \frac{((\frac{\kappa}{2})^2 + (2\pi\fij)^2)(2L\CT^3)}{\VDC^2 ~\frac{\SV^2}{2}\int\limits_{0}^{\infty}  df |\chi_{i,j}(f)|^2}}.
\label{eq:dcdu}
\end{equation}

With the amplitude of the driving noise set to $\VM = 2.7$~V$_\mathrm{rms}$, the corresponding spectral density is measured as $\SV^2=(3\pm0.1)\times10^{-10}$~V$^2$/Hz and we calculate $\int\limits_{0}^{\infty} df |\chi_{\mathrm{i,j}}(f)|^2$ numerically. Once we estimate $\frac{\partial \CC}{\partial u}$, Eq.~\ref{eq:go} gives us $\go$ for each observed mechanical mode (Fig.~3 of the main text). The error in this quantity reflects uncertainty in the cavity characterization ($A$, $\kappa$ and $\kappa_e$), the circuit model fit ($\CT$), the measurement of $\SV^2$, $P_\mathrm{side}$ and the fit of the mechanical sidebands ($\fij$ and $\Qij$).

 \begin{figure}[h]
\includegraphics[scale = 1.5]{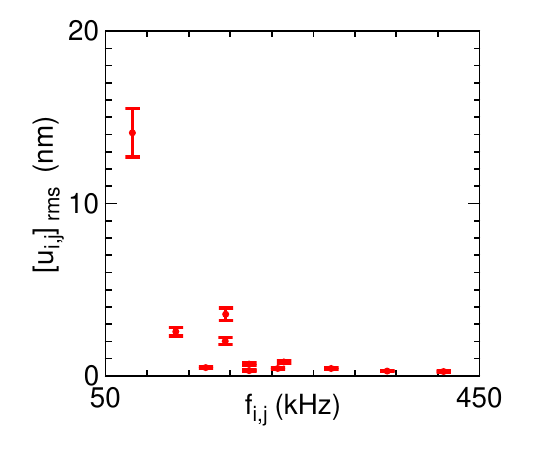}
\caption{\label{S3}
Extracted $[\delta u_{\mathrm{i,j}}]_\mathrm{rms}$ as a function of $\fij$ extracted from the measurements shown in Fig.~2 of the main text. 
}
\end{figure}

We also calculate the rms displacement $[\delta u_{\mathrm{i,j}}]_{\mathrm{rms}}$ corresponding to the mechanical sidebands in Fig.~3 of the main text using Eq.~\ref{eq:displacement}. For the fundamental mode, $[\delta u_{\mathrm{0}}]_{\mathrm{rms}}\sim 14$~nm (Fig.~\ref{S3}). For comparison, for a thermal state at 293~K $[\delta u_{\mathrm{0}}]_{\mathrm{rms}} = \uzp \sqrt{k_{\mathrm{B}}T/h\fo}\sim4$ pm. Therefore the white noise power spectrum of the electronic force is much larger than the power spectrum of the thermal noise $\VDC^2\SV^2 (\partial \CC/\partial u)^2 \gg 4\pi k_{\mathrm{B}}T\fij/\Qij$, justifying the approximation in Eq.~\ref{eq:F_el}.


\subsection{Mode profile correction to $\go$} 

The mode profile modifies $\frac{\partial\CC}{\partial u}$ and thus $\go$. For modes with $i \times j$ even, the sections of the membrane moving away from the antenna and the sections moving towards the antenna are equal, and therefore the net $\frac{\partial\CC}{\partial u}$ is close to zero, and $\go\sim0$. For $i \times j$ odd, there is always a section of the membrane which does not have a counterpart moving out of phase. Therefore, for odd mode profiles $\frac{\partial\CC}{\partial u}$, and thus $\go$, are reduced by a factor $1/(i\times j)$.




\section{Electromechanically induced transparency}

The transmission of our circuit in the presence of a weak probe tone at $\fP$ and a strong drive tone at $\fD$, as shown in the Fig.4(c-e) of the main text, is ~\cite{Weis2010supp, Aspelmeyer2014supp}:

\begin{equation}
|S_{21}(\deltaf)|=1-\frac{(1-i \mathcal{F}(\deltaf))\kappa_e}{2\pi i(\Delta+\deltaf)+\kappa/2+4\pi\Delta F(\deltaf)},
\label{eq:omit}
\end{equation} 
where $\deltaf=\fP-\fD$, $\Delta=\fD-\fC$ and  

\begin{equation}
 \mathcal{F}(\deltaf)=\frac{\hbar~\go^2~\chi_0(\deltaf)~\kappa_e ~S_{\mathrm{in}}^2}{u_\mathrm{ZP}^2(2\pi i(\deltaf-\Delta)+\kappa/2)((2\pi\Delta)^2+(\kappa/2)^2)}.
\end{equation}

We define $\chi_0$ as the mechanical susceptibility of the fundamental mode and $S_{\mathrm{in}}$ as the photon flux incident from the drive tone,

\begin{equation}
S_{\mathrm{in}}^2=\frac{\tilde{P}_{\mathrm{D}}}{2\pi\hbar\fD},
\end{equation} 
where $\tilde{P}_{\mathrm{D}}$ is the power of the drive tone ($P_{\mathrm{D}}$) having taken into account the overall insertion loss. For the measurements in Fig.~4 of the main text $P_{\mathrm{D}}=5$~dBm. 

Using the values of $\kappa/(2\pi)$ and $\kappa_e/(2\pi)$ extracted in section~\ref{sec:sectionI}, we fitted the curves in each panel of Fig.~4(c-e) of the main text with equation~\ref{eq:omit}. In this way, we obtained the reported values for $Q_0$, $f_0$ and $\go/2\pi$. The uncertainties in these quantities reflect the variance among the values obtained for each curve.

To estimate the amplitude of the membrane's motion we consider parametric amplification of the oscillator due to the beat frequency between the drive and probe tones~\cite{Weis2010supp}. For the OMIT measurement, the probe is 32.5~dB weaker than the drive (see Fig.~4 of the main text), hence the intracavity photon number is modulated by $N\approx 2\times10^{-32.5/20}n_\mathrm{C}$. Here we have neglected any difference due to the cavity resonance as both tones are well within the linewidth. The circulating photon number is estimated at $n_\mathrm{C} \approx 5.2 \times 10^{12}$ (at 5 dBm). For OMIT the beat frequency is at $\fo$, hence the oscillator sees the parametric force $F = \hbar g_0 N  \cos(2\pi \fo t)/\uzp$. This force results in a coherent amplitude  $u_{\mathrm{max}} = \frac{\hbar g_0 Q_0 N}{m \uzp (2\pi \fo)^2}\sim8.6$~nm. This is significantly above the thermal rms motion.



%


\begin{thebibliography}{23}%
\makeatletter
\providecommand \@ifxundefined [1]{%
 \@ifx{#1\undefined}
}%
\providecommand \@ifnum [1]{%
 \ifnum #1\expandafter \@firstoftwo
 \else \expandafter \@secondoftwo
 \fi
}%
\providecommand \@ifx [1]{%
 \ifx #1\expandafter \@firstoftwo
 \else \expandafter \@secondoftwo
 \fi
}%
\providecommand \natexlab [1]{#1}%
\providecommand \enquote  [1]{``#1''}%
\providecommand \bibnamefont  [1]{#1}%
\providecommand \bibfnamefont [1]{#1}%
\providecommand \citenamefont [1]{#1}%
\providecommand \href@noop [0]{\@secondoftwo}%
\providecommand \href [0]{\begingroup \@sanitize@url \@href}%
\providecommand \@href[1]{\@@startlink{#1}\@@href}%
\providecommand \@@href[1]{\endgroup#1\@@endlink}%
\providecommand \@sanitize@url [0]{\catcode `\\12\catcode `\$12\catcode
  `\&12\catcode `\#12\catcode `\^12\catcode `\_12\catcode `\%12\relax}%
\providecommand \@@startlink[1]{}%
\providecommand \@@endlink[0]{}%
\providecommand \url  [0]{\begingroup\@sanitize@url \@url }%
\providecommand \@url [1]{\endgroup\@href {#1}{\urlprefix }}%
\providecommand \urlprefix  [0]{URL }%
\providecommand \Eprint [0]{\href }%
\providecommand \doibase [0]{http://dx.doi.org/}%
\providecommand \selectlanguage [0]{\@gobble}%
\providecommand \bibinfo  [0]{\@secondoftwo}%
\providecommand \bibfield  [0]{\@secondoftwo}%
\providecommand \translation [1]{[#1]}%
\providecommand \BibitemOpen [0]{}%
\providecommand \bibitemStop [0]{}%
\providecommand \bibitemNoStop [0]{.\EOS\space}%
\providecommand \EOS [0]{\spacefactor3000\relax}%
\providecommand \BibitemShut  [1]{\csname bibitem#1\endcsname}%
\let\auto@bib@innerbib\@empty
\bibitem [{\citenamefont {Aspelmeyer}\ \emph {et~al.}(2014)\citenamefont
  {Aspelmeyer}, \citenamefont {Kippenberg},\ and\ \citenamefont
  {Marquard}}]{Aspelmeyer2014}%
  \BibitemOpen
  \bibfield  {author} {\bibinfo {author} {\bibfnamefont {M.}~\bibnamefont
  {Aspelmeyer}}, \bibinfo {author} {\bibfnamefont {T.~J.}\ \bibnamefont
  {Kippenberg}}, \ and\ \bibinfo {author} {\bibfnamefont {F.}~\bibnamefont
  {Marquard}},\ }\href {\doibase 10.1103/RevModPhys.86.1391} {\bibfield
  {journal} {\bibinfo  {journal} {Rev. Mod. Phys.}\ }\textbf {\bibinfo {volume}
  {86}},\ \bibinfo {pages} {1391} (\bibinfo {year} {2014})}\BibitemShut
  {NoStop}%
\bibitem [{\citenamefont {Purdy}\ \emph
  {et~al.}(2013{\natexlab{a}})\citenamefont {Purdy}, \citenamefont {Peterson},\
  and\ \citenamefont {Regal}}]{Purdy2013}%
  \BibitemOpen
  \bibfield  {author} {\bibinfo {author} {\bibfnamefont {T.~P.}\ \bibnamefont
  {Purdy}}, \bibinfo {author} {\bibfnamefont {R.~W.}\ \bibnamefont {Peterson}},
  \ and\ \bibinfo {author} {\bibfnamefont {C.~A.}\ \bibnamefont {Regal}},\
  }\href {\doibase 10.1126/science.1231692} {\bibfield  {journal} {\bibinfo
  {journal} {Science}\ }\textbf {\bibinfo {volume} {339}},\ \bibinfo {pages}
  {801} (\bibinfo {year} {2013}{\natexlab{a}})}\BibitemShut {NoStop}%
\bibitem [{\citenamefont {Purdy}\ \emph
  {et~al.}(2013{\natexlab{b}})\citenamefont {Purdy}, \citenamefont {Yu},
  \citenamefont {Peterson}, \citenamefont {Kampel},\ and\ \citenamefont
  {Regal}}]{Purdy2013a}%
  \BibitemOpen
  \bibfield  {author} {\bibinfo {author} {\bibfnamefont {T.~P.}\ \bibnamefont
  {Purdy}}, \bibinfo {author} {\bibfnamefont {P.~L.}\ \bibnamefont {Yu}},
  \bibinfo {author} {\bibfnamefont {R.~W.}\ \bibnamefont {Peterson}}, \bibinfo
  {author} {\bibfnamefont {N.~S.}\ \bibnamefont {Kampel}}, \ and\ \bibinfo
  {author} {\bibfnamefont {C.~A.}\ \bibnamefont {Regal}},\ }\href {\doibase
  10.1103/PhysRevX.3.031012} {\bibfield  {journal} {\bibinfo  {journal} {Phys.
  Rev. X}\ }\textbf {\bibinfo {volume} {3}},\ \bibinfo {pages} {031012}
  (\bibinfo {year} {2013}{\natexlab{b}})}\BibitemShut {NoStop}%
\bibitem [{\citenamefont {Andrews}\ \emph {et~al.}(2014)\citenamefont
  {Andrews}, \citenamefont {Peterson}, \citenamefont {Purdy}, \citenamefont
  {Cicak}, \citenamefont {Simmonds}, \citenamefont {Regal},\ and\ \citenamefont
  {Lehnert}}]{Andrews2014}%
  \BibitemOpen
  \bibfield  {author} {\bibinfo {author} {\bibfnamefont {R.~W.}\ \bibnamefont
  {Andrews}}, \bibinfo {author} {\bibfnamefont {R.~W.}\ \bibnamefont
  {Peterson}}, \bibinfo {author} {\bibfnamefont {T.~P.}\ \bibnamefont {Purdy}},
  \bibinfo {author} {\bibfnamefont {K.}~\bibnamefont {Cicak}}, \bibinfo
  {author} {\bibfnamefont {R.~W.}\ \bibnamefont {Simmonds}}, \bibinfo {author}
  {\bibfnamefont {C.~A.}\ \bibnamefont {Regal}}, \ and\ \bibinfo {author}
  {\bibfnamefont {K.~W.}\ \bibnamefont {Lehnert}},\ }\href {\doibase
  10.1038/nphys2911} {\bibfield  {journal} {\bibinfo  {journal} {Nature Phys.}\
  }\textbf {\bibinfo {volume} {10}},\ \bibinfo {pages} {321} (\bibinfo {year}
  {2014})}\BibitemShut {NoStop}%
\bibitem [{\citenamefont {Bagci}\ \emph {et~al.}(2014)\citenamefont {Bagci},
  \citenamefont {Simonsen}, \citenamefont {Schmid}, \citenamefont {Villanueva},
  \citenamefont {Zeuthen}, \citenamefont {Appel}, \citenamefont {Taylor},
  \citenamefont {S{\o}rensen}, \citenamefont {Usami}, \citenamefont
  {Schliesser},\ and\ \citenamefont {Polzik}}]{Bagci2014}%
  \BibitemOpen
  \bibfield  {author} {\bibinfo {author} {\bibfnamefont {T.}~\bibnamefont
  {Bagci}}, \bibinfo {author} {\bibfnamefont {A.}~\bibnamefont {Simonsen}},
  \bibinfo {author} {\bibfnamefont {S.}~\bibnamefont {Schmid}}, \bibinfo
  {author} {\bibfnamefont {L.~G.}\ \bibnamefont {Villanueva}}, \bibinfo
  {author} {\bibfnamefont {E.}~\bibnamefont {Zeuthen}}, \bibinfo {author}
  {\bibfnamefont {J.}~\bibnamefont {Appel}}, \bibinfo {author} {\bibfnamefont
  {J.~M.}\ \bibnamefont {Taylor}}, \bibinfo {author} {\bibfnamefont
  {A.}~\bibnamefont {S{\o}rensen}}, \bibinfo {author} {\bibfnamefont
  {K.}~\bibnamefont {Usami}}, \bibinfo {author} {\bibfnamefont
  {A.}~\bibnamefont {Schliesser}}, \ and\ \bibinfo {author} {\bibfnamefont
  {E.~S.}\ \bibnamefont {Polzik}},\ }\href {\doibase 10.1038/nature13029}
  {\bibfield  {journal} {\bibinfo  {journal} {Nature}\ }\textbf {\bibinfo
  {volume} {507}},\ \bibinfo {pages} {81} (\bibinfo {year} {2014})}\BibitemShut
  {NoStop}%
\bibitem [{\citenamefont {{Moaddel Haghighi}}\ \emph
  {et~al.}(2018)\citenamefont {{Moaddel Haghighi}}, \citenamefont {Malossi},
  \citenamefont {Natali}, \citenamefont {{Di Giuseppe}},\ and\ \citenamefont
  {Vitali}}]{Moaddel2018}%
  \BibitemOpen
  \bibfield  {author} {\bibinfo {author} {\bibfnamefont {I.}~\bibnamefont
  {{Moaddel Haghighi}}}, \bibinfo {author} {\bibfnamefont {N.}~\bibnamefont
  {Malossi}}, \bibinfo {author} {\bibfnamefont {R.}~\bibnamefont {Natali}},
  \bibinfo {author} {\bibfnamefont {G.}~\bibnamefont {{Di Giuseppe}}}, \ and\
  \bibinfo {author} {\bibfnamefont {D.}~\bibnamefont {Vitali}},\ }\href
  {\doibase 10.1103/PhysRevApplied.9.034031} {\bibfield  {journal} {\bibinfo
  {journal} {Phys. Rev. Appl.}\ }\textbf {\bibinfo {volume} {9}},\ \bibinfo
  {pages} {34031} (\bibinfo {year} {2018})}\BibitemShut {NoStop}%
\bibitem [{\citenamefont {Yuan}\ \emph
  {et~al.}(2015{\natexlab{a}})\citenamefont {Yuan}, \citenamefont {Singh},
  \citenamefont {Blanter},\ and\ \citenamefont {Steele}}]{Yuan2015}%
  \BibitemOpen
  \bibfield  {author} {\bibinfo {author} {\bibfnamefont {M.}~\bibnamefont
  {Yuan}}, \bibinfo {author} {\bibfnamefont {V.}~\bibnamefont {Singh}},
  \bibinfo {author} {\bibfnamefont {Y.~M.}\ \bibnamefont {Blanter}}, \ and\
  \bibinfo {author} {\bibfnamefont {G.~A.}\ \bibnamefont {Steele}},\ }\href
  {\doibase 10.1038/ncomms9491} {\bibfield  {journal} {\bibinfo  {journal}
  {Nat. Commun.}\ }\textbf {\bibinfo {volume} {6}},\ \bibinfo {pages} {8491}
  (\bibinfo {year} {2015}{\natexlab{a}})}\BibitemShut {NoStop}%
\bibitem [{\citenamefont {Noguchi}\ \emph {et~al.}(2016)\citenamefont
  {Noguchi}, \citenamefont {Yamazaki}, \citenamefont {Ataka}, \citenamefont
  {Fujita}, \citenamefont {Tabuchi}, \citenamefont {Ishikawa}, \citenamefont
  {Usami},\ and\ \citenamefont {Nakamura}}]{Nakamura2016}%
  \BibitemOpen
  \bibfield  {author} {\bibinfo {author} {\bibfnamefont {A.}~\bibnamefont
  {Noguchi}}, \bibinfo {author} {\bibfnamefont {R.}~\bibnamefont {Yamazaki}},
  \bibinfo {author} {\bibfnamefont {M.}~\bibnamefont {Ataka}}, \bibinfo
  {author} {\bibfnamefont {H.}~\bibnamefont {Fujita}}, \bibinfo {author}
  {\bibfnamefont {Y.}~\bibnamefont {Tabuchi}}, \bibinfo {author} {\bibfnamefont
  {T.}~\bibnamefont {Ishikawa}}, \bibinfo {author} {\bibfnamefont
  {K.}~\bibnamefont {Usami}}, \ and\ \bibinfo {author} {\bibfnamefont
  {Y.}~\bibnamefont {Nakamura}},\ }\href {\doibase
  10.1088/1367-2630/18/10/103036} {\bibfield  {journal} {\bibinfo  {journal}
  {New J. Phys.}\ }\textbf {\bibinfo {volume} {18}},\ \bibinfo {pages} {103036}
  (\bibinfo {year} {2016})}\BibitemShut {NoStop}%
\bibitem [{\citenamefont {Peterson}\ \emph {et~al.}(2016)\citenamefont
  {Peterson}, \citenamefont {Purdy}, \citenamefont {Kampel}, \citenamefont
  {Andrews}, \citenamefont {Yu}, \citenamefont {Lehnert},\ and\ \citenamefont
  {Regal}}]{Peterson2016}%
  \BibitemOpen
  \bibfield  {author} {\bibinfo {author} {\bibfnamefont {R.~W.}\ \bibnamefont
  {Peterson}}, \bibinfo {author} {\bibfnamefont {T.~P.}\ \bibnamefont {Purdy}},
  \bibinfo {author} {\bibfnamefont {N.~S.}\ \bibnamefont {Kampel}}, \bibinfo
  {author} {\bibfnamefont {R.~W.}\ \bibnamefont {Andrews}}, \bibinfo {author}
  {\bibfnamefont {P.~L.}\ \bibnamefont {Yu}}, \bibinfo {author} {\bibfnamefont
  {K.~W.}\ \bibnamefont {Lehnert}}, \ and\ \bibinfo {author} {\bibfnamefont
  {C.~A.}\ \bibnamefont {Regal}},\ }\href {\doibase
  10.1103/PhysRevLett.116.063601} {\bibfield  {journal} {\bibinfo  {journal}
  {Phys. Rev. Lett.}\ }\textbf {\bibinfo {volume} {116}},\ \bibinfo {pages}
  {063601} (\bibinfo {year} {2016})}\BibitemShut {NoStop}%
\bibitem [{\citenamefont {Fink}\ \emph {et~al.}(2016)\citenamefont {Fink},
  \citenamefont {Kalaee}, \citenamefont {Pitanti}, \citenamefont {Norte},
  \citenamefont {Heinzle}, \citenamefont {Davan\c{c}o}, \citenamefont
  {Srinivasan},\ and\ \citenamefont {Painter}}]{Fink2016}%
  \BibitemOpen
  \bibfield  {author} {\bibinfo {author} {\bibfnamefont {J.~M.}\ \bibnamefont
  {Fink}}, \bibinfo {author} {\bibfnamefont {M.}~\bibnamefont {Kalaee}},
  \bibinfo {author} {\bibfnamefont {A.}~\bibnamefont {Pitanti}}, \bibinfo
  {author} {\bibfnamefont {R.}~\bibnamefont {Norte}}, \bibinfo {author}
  {\bibfnamefont {L.}~\bibnamefont {Heinzle}}, \bibinfo {author} {\bibfnamefont
  {M.}~\bibnamefont {Davan\c{c}o}}, \bibinfo {author} {\bibfnamefont
  {K.}~\bibnamefont {Srinivasan}}, \ and\ \bibinfo {author} {\bibfnamefont
  {O.}~\bibnamefont {Painter}},\ }\href {\doibase 10.1038/ncomms12396}
  {\bibfield  {journal} {\bibinfo  {journal} {Nature Communs.}\ }\textbf
  {\bibinfo {volume} {7}},\ \bibinfo {pages} {1} (\bibinfo {year}
  {2016})}\BibitemShut {NoStop}%
\bibitem [{\citenamefont {Lehnert}(2014)}]{Lehnert2014}%
  \BibitemOpen
  \bibfield  {author} {\bibinfo {author} {\bibfnamefont {K.~W.}\ \bibnamefont
  {Lehnert}}\ }(\bibinfo  {publisher} {Springer Berlin Heidelberg},\ \bibinfo
  {address} {Berlin, Heidelberg},\ \bibinfo {year} {2014})\ pp.\ \bibinfo
  {pages} {233--252}\BibitemShut {NoStop}%
\bibitem [{\citenamefont {Ares}\ \emph
  {et~al.}(2016{\natexlab{a}})\citenamefont {Ares}, \citenamefont {Pei},
  \citenamefont {Mavalankar}, \citenamefont {Mergenthaler}, \citenamefont
  {Warner}, \citenamefont {Briggs},\ and\ \citenamefont {Laird}}]{Ares2016b}%
  \BibitemOpen
  \bibfield  {author} {\bibinfo {author} {\bibfnamefont {N.}~\bibnamefont
  {Ares}}, \bibinfo {author} {\bibfnamefont {T.}~\bibnamefont {Pei}}, \bibinfo
  {author} {\bibfnamefont {A.}~\bibnamefont {Mavalankar}}, \bibinfo {author}
  {\bibfnamefont {M.}~\bibnamefont {Mergenthaler}}, \bibinfo {author}
  {\bibfnamefont {J.~H.}\ \bibnamefont {Warner}}, \bibinfo {author}
  {\bibfnamefont {G.~A.~D.}\ \bibnamefont {Briggs}}, \ and\ \bibinfo {author}
  {\bibfnamefont {E.~A.}\ \bibnamefont {Laird}},\ }\href {\doibase
  10.1103/PhysRevLett.117.170801} {\bibfield  {journal} {\bibinfo  {journal}
  {Phys. Rev. Lett.}\ }\textbf {\bibinfo {volume} {117}},\ \bibinfo {pages}
  {170801} (\bibinfo {year} {2016}{\natexlab{a}})}\BibitemShut {NoStop}%
\bibitem [{\citenamefont {Brown}\ \emph {et~al.}(2007)\citenamefont {Brown},
  \citenamefont {Britton}, \citenamefont {Epstein}, \citenamefont {Chiaverini},
  \citenamefont {Leibfried},\ and\ \citenamefont {Wineland}}]{Brown2007}%
  \BibitemOpen
  \bibfield  {author} {\bibinfo {author} {\bibfnamefont {K.~R.}\ \bibnamefont
  {Brown}}, \bibinfo {author} {\bibfnamefont {J.}~\bibnamefont {Britton}},
  \bibinfo {author} {\bibfnamefont {R.}~\bibnamefont {Epstein}}, \bibinfo
  {author} {\bibfnamefont {J.}~\bibnamefont {Chiaverini}}, \bibinfo {author}
  {\bibfnamefont {D.}~\bibnamefont {Leibfried}}, \ and\ \bibinfo {author}
  {\bibfnamefont {D.~J.}\ \bibnamefont {Wineland}},\ }\href {\doibase
  10.1103/PhysRevLett.99.137205} {\bibfield  {journal} {\bibinfo  {journal}
  {Physical review letters}\ }\textbf {\bibinfo {volume} {99}},\ \bibinfo
  {pages} {137205} (\bibinfo {year} {2007})}\BibitemShut {NoStop}%
\bibitem [{\citenamefont {Faust}\ \emph {et~al.}(2012)\citenamefont {Faust},
  \citenamefont {Krenn}, \citenamefont {Manus}, \citenamefont {Kotthaus},\ and\
  \citenamefont {Weig}}]{Faust2012}%
  \BibitemOpen
  \bibfield  {author} {\bibinfo {author} {\bibfnamefont {T.}~\bibnamefont
  {Faust}}, \bibinfo {author} {\bibfnamefont {P.}~\bibnamefont {Krenn}},
  \bibinfo {author} {\bibfnamefont {S.}~\bibnamefont {Manus}}, \bibinfo
  {author} {\bibfnamefont {J.~P.}\ \bibnamefont {Kotthaus}}, \ and\ \bibinfo
  {author} {\bibfnamefont {E.~M.}\ \bibnamefont {Weig}},\ }\href {\doibase
  10.1038/ncomms1723} {\bibfield  {journal} {\bibinfo  {journal} {Nature
  communications}\ }\textbf {\bibinfo {volume} {3}},\ \bibinfo {pages} {728}
  (\bibinfo {year} {2012})}\BibitemShut {NoStop}%
\bibitem [{\citenamefont {Weis}\ \emph {et~al.}(2010)\citenamefont {Weis},
  \citenamefont {Rivi{\`{e}}re}, \citenamefont {Del{\'{e}}glise}, \citenamefont
  {Gavartin}, \citenamefont {Arcizet}, \citenamefont {Schliesser},\ and\
  \citenamefont {Kippenberg}}]{Weis2010}%
  \BibitemOpen
  \bibfield  {author} {\bibinfo {author} {\bibfnamefont {S.}~\bibnamefont
  {Weis}}, \bibinfo {author} {\bibfnamefont {R.}~\bibnamefont {Rivi{\`{e}}re}},
  \bibinfo {author} {\bibfnamefont {S.}~\bibnamefont {Del{\'{e}}glise}},
  \bibinfo {author} {\bibfnamefont {E.}~\bibnamefont {Gavartin}}, \bibinfo
  {author} {\bibfnamefont {O.}~\bibnamefont {Arcizet}}, \bibinfo {author}
  {\bibfnamefont {A.}~\bibnamefont {Schliesser}}, \ and\ \bibinfo {author}
  {\bibfnamefont {T.~J.}\ \bibnamefont {Kippenberg}},\ }\href {\doibase
  10.1126/science.1195596} {\bibfield  {journal} {\bibinfo  {journal}
  {Science}\ }\textbf {\bibinfo {volume} {330}},\ \bibinfo {pages} {1520}
  (\bibinfo {year} {2010})}\BibitemShut {NoStop}%
\bibitem [{\citenamefont {Safavi-Naeini}\ \emph {et~al.}(2011)\citenamefont
  {Safavi-Naeini}, \citenamefont {Alegre}, \citenamefont {Chan}, \citenamefont
  {Eichenfield}, \citenamefont {Winger}, \citenamefont {Lin}, \citenamefont
  {Hill}, \citenamefont {Chang},\ and\ \citenamefont
  {Painter}}]{Safavi-Naeini2011}%
  \BibitemOpen
  \bibfield  {author} {\bibinfo {author} {\bibfnamefont {A.~H.}\ \bibnamefont
  {Safavi-Naeini}}, \bibinfo {author} {\bibfnamefont {T.~P.~M.}\ \bibnamefont
  {Alegre}}, \bibinfo {author} {\bibfnamefont {J.}~\bibnamefont {Chan}},
  \bibinfo {author} {\bibfnamefont {M.}~\bibnamefont {Eichenfield}}, \bibinfo
  {author} {\bibfnamefont {M.}~\bibnamefont {Winger}}, \bibinfo {author}
  {\bibfnamefont {Q.}~\bibnamefont {Lin}}, \bibinfo {author} {\bibfnamefont
  {J.~T.}\ \bibnamefont {Hill}}, \bibinfo {author} {\bibfnamefont {D.~E.}\
  \bibnamefont {Chang}}, \ and\ \bibinfo {author} {\bibfnamefont
  {O.}~\bibnamefont {Painter}},\ }\href {\doibase 10.1038/nature09933}
  {\bibfield  {journal} {\bibinfo  {journal} {Nature}\ }\textbf {\bibinfo
  {volume} {472}},\ \bibinfo {pages} {69} (\bibinfo {year} {2011})}\BibitemShut
  {NoStop}%
\bibitem [{\citenamefont {Ojanen}\ and\ \citenamefont
  {B{\o}rkje}(2014)}]{Ojanen}%
  \BibitemOpen
  \bibfield  {author} {\bibinfo {author} {\bibfnamefont {T.}~\bibnamefont
  {Ojanen}}\ and\ \bibinfo {author} {\bibfnamefont {K.}~\bibnamefont
  {B{\o}rkje}},\ }\href {\doibase 10.1103/PhysRevA.90.013824} {\bibfield
  {journal} {\bibinfo  {journal} {Phys. Rev. A}\ }\textbf {\bibinfo {volume}
  {90}},\ \bibinfo {pages} {013824} (\bibinfo {year} {2014})}\BibitemShut
  {NoStop}%
\bibitem [{\citenamefont {Yong-Chun}\ \emph {et~al.}(2015)\citenamefont
  {Yong-Chun}, \citenamefont {Yun-Feng}, \citenamefont {Luan},\ and\
  \citenamefont {Wong}}]{Yong-Chun2015}%
  \BibitemOpen
  \bibfield  {author} {\bibinfo {author} {\bibfnamefont {L.}~\bibnamefont
  {Yong-Chun}}, \bibinfo {author} {\bibfnamefont {X.}~\bibnamefont {Yun-Feng}},
  \bibinfo {author} {\bibfnamefont {X.-S.}\ \bibnamefont {Luan}}, \ and\
  \bibinfo {author} {\bibfnamefont {C.~W.}\ \bibnamefont {Wong}},\ }\href
  {\doibase 10.1007/s11433-014-5635-6} {\bibfield  {journal} {\bibinfo
  {journal} {Sci. China Phys. Mech.}\ }\textbf {\bibinfo {volume} {58}},\
  \bibinfo {pages} {1} (\bibinfo {year} {2015})}\BibitemShut {NoStop}%
\bibitem [{\citenamefont {Ares}\ \emph
  {et~al.}(2016{\natexlab{b}})\citenamefont {Ares}, \citenamefont {Schupp},
  \citenamefont {Mavalankar}, \citenamefont {Rogers}, \citenamefont
  {Griffiths}, \citenamefont {Jones}, \citenamefont {Farrer}, \citenamefont
  {Ritchie}, \citenamefont {Smith}, \citenamefont {Cottet}, \citenamefont
  {Briggs},\ and\ \citenamefont {Laird}}]{Ares2016}%
  \BibitemOpen
  \bibfield  {author} {\bibinfo {author} {\bibfnamefont {N.}~\bibnamefont
  {Ares}}, \bibinfo {author} {\bibfnamefont {F.~J.}\ \bibnamefont {Schupp}},
  \bibinfo {author} {\bibfnamefont {A.}~\bibnamefont {Mavalankar}}, \bibinfo
  {author} {\bibfnamefont {G.}~\bibnamefont {Rogers}}, \bibinfo {author}
  {\bibfnamefont {J.}~\bibnamefont {Griffiths}}, \bibinfo {author}
  {\bibfnamefont {G.~A.~C.}\ \bibnamefont {Jones}}, \bibinfo {author}
  {\bibfnamefont {I.}~\bibnamefont {Farrer}}, \bibinfo {author} {\bibfnamefont
  {D.~A.}\ \bibnamefont {Ritchie}}, \bibinfo {author} {\bibfnamefont {C.~G.}\
  \bibnamefont {Smith}}, \bibinfo {author} {\bibfnamefont {A.}~\bibnamefont
  {Cottet}}, \bibinfo {author} {\bibfnamefont {G.~A.~D.}\ \bibnamefont
  {Briggs}}, \ and\ \bibinfo {author} {\bibfnamefont {E.~A.}\ \bibnamefont
  {Laird}},\ }\href {\doibase 10.1103/PhysRevApplied.5.034011} {\bibfield
  {journal} {\bibinfo  {journal} {Phys. Rev. Appl.}\ }\textbf {\bibinfo
  {volume} {5}},\ \bibinfo {pages} {34011} (\bibinfo {year}
  {2016}{\natexlab{b}})}\BibitemShut {NoStop}%
\bibitem [{Sup()}]{Supp}%
  \BibitemOpen
  \href@noop {} {}\bibinfo {note} {See Supplemental Material for details of the
  circuit simulation, mechanical quality factors and estimation of the
  single-photon coupling from noise and OMIT measurements.}\BibitemShut {Stop}%
\bibitem [{\citenamefont {Schneider}\ \emph {et~al.}(2014)\citenamefont
  {Schneider}, \citenamefont {Singh}, \citenamefont {Venstra}, \citenamefont
  {Meerwaldt},\ and\ \citenamefont {Steele}}]{Schneider2014}%
  \BibitemOpen
  \bibfield  {author} {\bibinfo {author} {\bibfnamefont {B.~H.}\ \bibnamefont
  {Schneider}}, \bibinfo {author} {\bibfnamefont {V.}~\bibnamefont {Singh}},
  \bibinfo {author} {\bibfnamefont {W.~J.}\ \bibnamefont {Venstra}}, \bibinfo
  {author} {\bibfnamefont {H.~B.}\ \bibnamefont {Meerwaldt}}, \ and\ \bibinfo
  {author} {\bibfnamefont {G.~A.}\ \bibnamefont {Steele}},\ }\href {\doibase
  10.1038/ncomms6819} {\bibfield  {journal} {\bibinfo  {journal} {Nat.
  Commun.}\ }\textbf {\bibinfo {volume} {5}},\ \bibinfo {pages} {5819}
  (\bibinfo {year} {2014})}\BibitemShut {NoStop}%
\bibitem [{\citenamefont {Zwickl}\ \emph {et~al.}(2008)\citenamefont {Zwickl},
  \citenamefont {Shanks}, \citenamefont {Jayich}, \citenamefont {Yang},
  \citenamefont {{Bleszynski Jayich}}, \citenamefont {Thompson},\ and\
  \citenamefont {Harris}}]{Zwickl2008}%
  \BibitemOpen
  \bibfield  {author} {\bibinfo {author} {\bibfnamefont {B.~M.}\ \bibnamefont
  {Zwickl}}, \bibinfo {author} {\bibfnamefont {W.~E.}\ \bibnamefont {Shanks}},
  \bibinfo {author} {\bibfnamefont {A.~M.}\ \bibnamefont {Jayich}}, \bibinfo
  {author} {\bibfnamefont {C.}~\bibnamefont {Yang}}, \bibinfo {author}
  {\bibfnamefont {A.~C.}\ \bibnamefont {{Bleszynski Jayich}}}, \bibinfo
  {author} {\bibfnamefont {J.~D.}\ \bibnamefont {Thompson}}, \ and\ \bibinfo
  {author} {\bibfnamefont {J.~G.~E.}\ \bibnamefont {Harris}},\ }\href {\doibase
  10.1063/1.2884191} {\bibfield  {journal} {\bibinfo  {journal} {Appl. Phys.
  Lett.}\ }\textbf {\bibinfo {volume} {92}},\ \bibinfo {pages} {103125}
  (\bibinfo {year} {2008})}\BibitemShut {NoStop}%
\bibitem [{\citenamefont {Yuan}\ \emph
  {et~al.}(2015{\natexlab{b}})\citenamefont {Yuan}, \citenamefont {Cohen},\
  and\ \citenamefont {Steele}}]{Yuan2015a}%
  \BibitemOpen
  \bibfield  {author} {\bibinfo {author} {\bibfnamefont {M.}~\bibnamefont
  {Yuan}}, \bibinfo {author} {\bibfnamefont {M.~A.}\ \bibnamefont {Cohen}}, \
  and\ \bibinfo {author} {\bibfnamefont {G.~A.}\ \bibnamefont {Steele}},\
  }\href {\doibase 10.1063/1.4938747} {\bibfield  {journal} {\bibinfo
  {journal} {Appl. Phys. Lett.}\ }\textbf {\bibinfo {volume} {107}},\ \bibinfo
  {pages} {263501} (\bibinfo {year} {2015}{\natexlab{b}})}\BibitemShut
  {NoStop}%
\end{thebibliography}

\begin{thebibliography}{8}%
\makeatletter
\providecommand \@ifxundefined [1]{%
 \@ifx{#1\undefined}
}%
\providecommand \@ifnum [1]{%
 \ifnum #1\expandafter \@firstoftwo
 \else \expandafter \@secondoftwo
 \fi
}%
\providecommand \@ifx [1]{%
 \ifx #1\expandafter \@firstoftwo
 \else \expandafter \@secondoftwo
 \fi
}%
\providecommand \natexlab [1]{#1}%
\providecommand \enquote  [1]{``#1''}%
\providecommand \bibnamefont  [1]{#1}%
\providecommand \bibfnamefont [1]{#1}%
\providecommand \citenamefont [1]{#1}%
\providecommand \href@noop [0]{\@secondoftwo}%
\providecommand \href [0]{\begingroup \@sanitize@url \@href}%
\providecommand \@href[1]{\@@startlink{#1}\@@href}%
\providecommand \@@href[1]{\endgroup#1\@@endlink}%
\providecommand \@sanitize@url [0]{\catcode `\\12\catcode `\$12\catcode
  `\&12\catcode `\#12\catcode `\^12\catcode `\_12\catcode `\%12\relax}%
\providecommand \@@startlink[1]{}%
\providecommand \@@endlink[0]{}%
\providecommand \url  [0]{\begingroup\@sanitize@url \@url }%
\providecommand \@url [1]{\endgroup\@href {#1}{\urlprefix }}%
\providecommand \urlprefix  [0]{URL }%
\providecommand \Eprint [0]{\href }%
\providecommand \doibase [0]{http://dx.doi.org/}%
\providecommand \selectlanguage [0]{\@gobble}%
\providecommand \bibinfo  [0]{\@secondoftwo}%
\providecommand \bibfield  [0]{\@secondoftwo}%
\providecommand \translation [1]{[#1]}%
\providecommand \BibitemOpen [0]{}%
\providecommand \bibitemStop [0]{}%
\providecommand \bibitemNoStop [0]{.\EOS\space}%
\providecommand \EOS [0]{\spacefactor3000\relax}%
\providecommand \BibitemShut  [1]{\csname bibitem#1\endcsname}%
\let\auto@bib@innerbib\@empty

\bibitem [{\citenamefont {Singh}\ \emph {et~al.}(2014)\citenamefont {Singh},
  \citenamefont {Bosman}, \citenamefont {Schneider}, \citenamefont {Blanter},
  \citenamefont {Castellanos-Gomez},\ and\ \citenamefont {Steele}}]{Singh2014}%
  \BibitemOpen
  \bibfield  {author} {\bibinfo {author} {\bibfnamefont {V.}~\bibnamefont
  {Singh}}, \bibinfo {author} {\bibfnamefont {S.~J.}\ \bibnamefont {Bosman}},
  \bibinfo {author} {\bibfnamefont {B.~H.}\ \bibnamefont {Schneider}}, \bibinfo
  {author} {\bibfnamefont {Y.~M.}\ \bibnamefont {Blanter}}, \bibinfo {author}
  {\bibfnamefont {A.}~\bibnamefont {Castellanos-Gomez}}, \ and\ \bibinfo
  {author} {\bibfnamefont {G.~A.}\ \bibnamefont {Steele}},\ }\href {\doibase
  10.1038/nnano.2014.168} {\bibfield  {journal} {\bibinfo  {journal} {Nat.
  Nanotechnol.}\ }\textbf {\bibinfo {volume} {9}},\ \bibinfo {pages} {820}
  (\bibinfo {year} {2014})}\BibitemShut {NoStop}%
\bibitem [{\citenamefont {Lacy}(2011)}]{Lacy2011}%
  \BibitemOpen
  \bibfield  {author} {\bibinfo {author} {\bibfnamefont {F.}~\bibnamefont
  {Lacy}},\ }\href {\doibase 10.1186/1556-276X-6-636} {\bibfield  {journal}
  {\bibinfo  {journal} {Nanoscale Res. Lett.}\ }\textbf {\bibinfo {volume}
  {6}},\ \bibinfo {pages} {1} (\bibinfo {year} {2011})}\BibitemShut {NoStop}%
\bibitem [{\citenamefont {Ares}\ \emph {et~al.}(2016)\citenamefont {Ares},
  \citenamefont {Schupp}, \citenamefont {Mavalankar}, \citenamefont {Rogers},
  \citenamefont {Griffiths}, \citenamefont {Jones}, \citenamefont {Farrer},
  \citenamefont {Ritchie}, \citenamefont {Smith}, \citenamefont {Cottet},
  \citenamefont {Briggs},\ and\ \citenamefont {Laird}}]{Ares2016supp}%
  \BibitemOpen
  \bibfield  {author} {\bibinfo {author} {\bibfnamefont {N.}~\bibnamefont
  {Ares}}, \bibinfo {author} {\bibfnamefont {F.~J.}\ \bibnamefont {Schupp}},
  \bibinfo {author} {\bibfnamefont {A.}~\bibnamefont {Mavalankar}}, \bibinfo
  {author} {\bibfnamefont {G.}~\bibnamefont {Rogers}}, \bibinfo {author}
  {\bibfnamefont {J.}~\bibnamefont {Griffiths}}, \bibinfo {author}
  {\bibfnamefont {G.~A.~C.}\ \bibnamefont {Jones}}, \bibinfo {author}
  {\bibfnamefont {I.}~\bibnamefont {Farrer}}, \bibinfo {author} {\bibfnamefont
  {D.~A.}\ \bibnamefont {Ritchie}}, \bibinfo {author} {\bibfnamefont {C.~G.}\
  \bibnamefont {Smith}}, \bibinfo {author} {\bibfnamefont {A.}~\bibnamefont
  {Cottet}}, \bibinfo {author} {\bibfnamefont {G.~A.~D.}\ \bibnamefont
  {Briggs}}, \ and\ \bibinfo {author} {\bibfnamefont {E.~A.}\ \bibnamefont
  {Laird}},\ }\href {\doibase 10.1103/PhysRevApplied.5.034011} {\bibfield
  {journal} {\bibinfo  {journal} {Phys. Rev. Appl.}\ }\textbf {\bibinfo
  {volume} {5}},\ \bibinfo {pages} {34011} (\bibinfo {year}
  {2016})}\BibitemShut {NoStop}%
\bibitem [{\citenamefont {Poot}\ and\ \citenamefont {van~der
  Zant}(2012)}]{Poot2012}%
  \BibitemOpen
  \bibfield  {author} {\bibinfo {author} {\bibfnamefont {M.}~\bibnamefont
  {Poot}}\ and\ \bibinfo {author} {\bibfnamefont {H.~S.~J.}\ \bibnamefont
  {van~der Zant}},\ }\href {\doibase 10.1016/j.physrep.2011.12.004} {\bibfield
  {journal} {\bibinfo  {journal} {Phys. Rep.}\ }\textbf {\bibinfo {volume}
  {511}},\ \bibinfo {pages} {273} (\bibinfo {year} {2012})}\BibitemShut
  {NoStop}%
\bibitem [{\citenamefont {Yuan}\ \emph {et~al.}(2015)\citenamefont {Yuan},
  \citenamefont {Singh}, \citenamefont {Blanter},\ and\ \citenamefont
  {Steele}}]{Yuan2015supp}%
  \BibitemOpen
  \bibfield  {author} {\bibinfo {author} {\bibfnamefont {M.}~\bibnamefont
  {Yuan}}, \bibinfo {author} {\bibfnamefont {V.}~\bibnamefont {Singh}},
  \bibinfo {author} {\bibfnamefont {Y.~M.}\ \bibnamefont {Blanter}}, \ and\
  \bibinfo {author} {\bibfnamefont {G.~A.}\ \bibnamefont {Steele}},\ }\href
  {\doibase 10.1038/ncomms9491} {\bibfield  {journal} {\bibinfo  {journal}
  {Nat. Commun.}\ }\textbf {\bibinfo {volume} {6}},\ \bibinfo {pages} {8491}
  (\bibinfo {year} {2015})}\BibitemShut {NoStop}%
\bibitem [{\citenamefont {Aspelmeyer}\ \emph {et~al.}(2014)\citenamefont
  {Aspelmeyer}, \citenamefont {Kippenberg},\ and\ \citenamefont
  {Marquard}}]{Aspelmeyer2014supp}%
  \BibitemOpen
  \bibfield  {author} {\bibinfo {author} {\bibfnamefont {M.}~\bibnamefont
  {Aspelmeyer}}, \bibinfo {author} {\bibfnamefont {T.~J.}\ \bibnamefont
  {Kippenberg}}, \ and\ \bibinfo {author} {\bibfnamefont {F.}~\bibnamefont
  {Marquard}},\ }\href {\doibase 10.1103/RevModPhys.86.1391} {\bibfield
  {journal} {\bibinfo  {journal} {Rev. Mod. Phys.}\ }\textbf {\bibinfo {volume}
  {86}},\ \bibinfo {pages} {1391} (\bibinfo {year} {2014})}\BibitemShut
  {NoStop}%
\bibitem [{\citenamefont {Lehnert}(2014)}]{Lehnert2014supp}%
  \BibitemOpen
  \bibfield  {author} {\bibinfo {author} {\bibfnamefont {K.~W.}\ \bibnamefont
  {Lehnert}}\ }(\bibinfo  {publisher} {Springer Berlin Heidelberg},\ \bibinfo
  {address} {Berlin, Heidelberg},\ \bibinfo {year} {2014})\ pp.\ \bibinfo
  {pages} {233--252}\BibitemShut {NoStop}%
\bibitem [{\citenamefont {Weis}\ \emph {et~al.}(2010)\citenamefont {Weis},
  \citenamefont {Rivi{\`{e}}re}, \citenamefont {Del{\'{e}}glise}, \citenamefont
  {Gavartin}, \citenamefont {Arcizet}, \citenamefont {Schliesser},\ and\
  \citenamefont {Kippenberg}}]{Weis2010supp}%
  \BibitemOpen
  \bibfield  {author} {\bibinfo {author} {\bibfnamefont {S.}~\bibnamefont
  {Weis}}, \bibinfo {author} {\bibfnamefont {R.}~\bibnamefont {Rivi{\`{e}}re}},
  \bibinfo {author} {\bibfnamefont {S.}~\bibnamefont {Del{\'{e}}glise}},
  \bibinfo {author} {\bibfnamefont {E.}~\bibnamefont {Gavartin}}, \bibinfo
  {author} {\bibfnamefont {O.}~\bibnamefont {Arcizet}}, \bibinfo {author}
  {\bibfnamefont {A.}~\bibnamefont {Schliesser}}, \ and\ \bibinfo {author}
  {\bibfnamefont {T.~J.}\ \bibnamefont {Kippenberg}},\ }\href {\doibase
  10.1126/science.1195596} {\bibfield  {journal} {\bibinfo  {journal}
  {Science}\ }\textbf {\bibinfo {volume} {330}},\ \bibinfo {pages} {1520}
  (\bibinfo {year} {2010})}\BibitemShut {NoStop}%
\end{thebibliography}
\end{document}